\pdfoutput=1
\documentclass[prd,aps,superscriptaddress,amsmath,amssymb,twocolumn,nobibnotes,nofootinbib,10pt,groupedaddress,longbibliography]{revtex4-2}

\usepackage{braket,verbatim,bm,bbold,amsmath,amssymb,epsfig,xcolor,float}
\definecolor{redCB}{RGB}{150,50,60}  
\definecolor{dteal}{RGB}{7,94,84}  
\definecolor{blueCB}{RGB}{108,144,170}
\usepackage[colorlinks=true,    	
			pdfstartview=FitV,
            bookmarks=false,
            citecolor=dteal,
            linkcolor=dteal,
           linkcolor=dteal,
            hyperfootnotes=true,
            linktoc=page,
            urlcolor=dteal]{hyperref}
\hypersetup{breaklinks=true}
\usepackage{soul}
\usepackage{tikzlings}
\usepackage{tikzducks}
\usepackage{flushend}
\graphicspath{{./figs/}}

\newcommand{\nc}{\newcommand}

\def\M{\mathcal{M}}
\def\({\left(}
\def\){\right)}
\def\[{\left[}
\def\]{\right]}

\def\o{\omega}

\def\D{\mathcal{D}}
\nc{\ir}{\mathrm{i}}
\nc{\dd}{\mathrm{d}} 
\nc{\eE}{\mathrm{e}}
\nc{\tr}{\text{Tr}}
\nc{\Tr}{\text{Tr}}
\nc{\id}{\mathbb{I}}
\nc{\Z}{\mathcal{Z}}
\nc{\E}{\mathcal{E}}
\nc{\F}{\mathcal{F}}
\nc{\Om}{\Omega}
\nc{\N}{\mathcal{N}}
\nc{\spp}{\hspace{1pt}}
\nc{\spm}{\hspace{-1pt}}
\nc{\lb}{\label} 
\nc{\nn}{\nonumber} 
\nc{\ra}{\rangle} 
\nc{\la}{\langle} 
\nc{\Nn}{\mathcal{N}} 
\nc{\sigb}{\boldsymbol{\sigma_{\text{bd}}}}
\nc{\Hh}{\mathcal{H}}
\nc{\HA}{\mathcal{H}_A}
\nc{\HB}{\mathcal{H}_B}
\nc{\HC}{\mathcal{H}_C}
\nc{\ta}{\tilde{a}}
\nc{\eps}{\epsilon}
\nc{\Q}{\mathcal{Q}}
\nc{\T}{\mathcal{T}}
\nc{\Oo}{\mathcal{O}}


\def\bea#1\eea{\begin{align}#1\end{align}}
\def\bes#1\ees{\begin{equation}\begin{split}#1\end{split}\end{equation}}

\makeatletter                                                                 
    \def\footnoterule{\kern -6\p@        
  \hrule \@width 0.5in \kern 5.7\p@}  
\makeatother

\begin{document}

\title{Genuine multientropy, dihedral invariants and Lifshitz theory}

\author{Cl\'ement Berthière}
\email{clement.berthiere@irsamc.ups-tlse.fr}
\affiliation{\vspace{3pt}Laboratoire de Physique Théorique, CNRS, Université de Toulouse, France}

\author{Paul Gaudin}
\email{paul.gaudin@irsamc.ups-tlse.fr}
\affiliation{\vspace{3pt}Laboratoire de Physique Théorique, CNRS, Université de Toulouse, France}

\date{\today}

\begin{abstract}
Multi-invariants are local-unitary invariants of state replicas introduced as potential probes of multipartite entanglement and correlations in quantum many-body systems. In this paper, we investigate two multi-invariants for tripartite pure states, namely multientropy and dihedral invariant. We compute the (genuine) multientropy for Lifshitz groundstates, and obtain its analytical continuation to noninteger values of Rényi index. We show that the genuine multientropy can be expressed in terms of mutual information and logarithmic negativity, a relation that also holds for stabilizer states. For general tripartite pure states, we demonstrate that dihedral invariants are related to Rényi reflected entropies. In particular, we show that the dihedral permutations of replicas are equivalent to the reflected construction, or alternatively to the realignment of density matrices.

\end{abstract}

\maketitle
\makeatletter

\def\l@subsubsection#1#2{}
\makeatother


\section{Introduction}

Bipartite entanglement has played a central role in the study of quantum matter \cite{Amico:2007ag,Laflorencie:2015eck}. In particular, its measures have been remarkably successful in diagnosing universal scaling laws at criticality \cite{Vidal:2002rm,Calabrese:2004eu,2006PhRvL..96j0603L,Eisert:2008ur,2009JPhA...42X4009A,Bueno:2015rda,Fursaev:2016inw,Casini:2016fgb,Berthiere:2018ouo,Berthiere:2019lks,Berthiere:2021nkv}, characterizing topological order \cite{Kitaev:2005dm,2006PhRvL..96k0405L,2015arXiv150802595Z}, and even offering a unifying language across fields ranging from condensed matter physics to quantum gravity \cite{Rangamani:2016dms,Takayanagi:2025ula}. Yet, bipartite entanglement is only the tip of the iceberg---it cannot fully characterize quantum systems composed of many parties. Understanding the multipartite entanglement structure of many-body quantum states is thus fundamental to discovering and diagnosing novel phases of matter and phenomena, see, e.g., \cite{Akers:2019gcv,Zou:2020bly,Hayden:2021gno,Liu:2021ctk,tam2022topological,Parez:2022ind,Parez:2022uva,Liu:2023pdz,Berthiere:2023bwn,Parez:2024zbz,Wen:2024pcj,Berthiere:2024sio} for recent studies. 

In this direction, the notion of multi-invariants \cite{Gadde:2024taa} has recently been introduced as potential probes of multipartite entanglement. These are local unitary invariants built from copies of the density matrix of a quantum multipartite state, and naturally generalize bipartite Rényi entanglement entropy by extending the group symmetry of replicas.
Focusing on tripartite systems, we investigate the multientropy \cite{Gadde:2022cqi,Penington:2022dhr} and dihedral invariants \cite{Gadde:2024taa}, which stand out as tractable multi-invariants. 

The purpose of this paper is twofold. First, we perform the calculation of multientropy for Lifshitz groundstates, which allow for a full analytical treatment in quantum field theory. In particular, we are interested in the genuine multientropy, defined as the difference between the multientropy and the averaged sum of Rényi entanglement entropies, as it is UV finite and expected to characterize the amount of genuine tripartite entanglement. Our second goal is to show that, for general tripartite pure states, dihedral invariants are directly related to a well-known quantity---the Rényi reflected entropy.

\medskip
\textit{Organization of the paper} --- In Section~\ref{sec:def}, we review the definition of (genuine) multientropy and dihedral invariants, and provide some background on Lifshitz theories and the peculiar structure of their groundstates. We compute the genuine multientropy for Lifshitz groundstates in Section~\ref{sec:GM}, where two different tripartitions are considered. We find that it can be expressed in terms of mutual information and logarithmic negativity, and discuss this relation. The behavior of genuine multientropy is also compared to that of the Markov gap---a probe of tripartite entanglement. 
In Section~\ref{sec:dihedral}, we demonstrate that dihedral invariants are exactly Rényi reflected entropies, for general tripartite pure states.
We conclude in Section~\ref{sec:conclu} with a summary of our main results, and give an outlook on future studies. Finally, technical details and further calculations can be found in the three appendices that complete this work.

\section{Preliminaries and main results}\lb{sec:def}

\subsection{Preliminaries}

\textit{Genuine multientropy} --- 
Multientropy \cite{Gadde:2022cqi,Penington:2022dhr} is a quantity proposed to capture multipartite entanglement. For a tripartite pure quantum state $|\Psi\ra$ on $\HA\otimes\HB\otimes\HC$, the Rényi multientropy is defined via the replica trick
\bea
\begin{split}\lb{def_multi}
&S^{(3)}_n(A:B:C) = \frac{1}{1-n}\frac1n \log\frac{\Z^{(3)}_n}{(\Z_1^{(3)})^{n^2}},\\[1ex]
& \Z^{(3)}_n=\la\Psi|^{\otimes n^2}(\pi_A\otimes\pi_B\otimes\pi_C)|\Psi\ra^{\otimes n^2}.
\end{split}
\eea
The permutation operator $\pi_K$ is in the group $\mathbb{S}_{n^2}$ and permutes the $n^2$ copies of the corresponding subsystem $K=\{A,B,C\}$, explicitly
\bea
&\pi_A :\; (1,2,\dots\hspace{-.5pt},n)(n+1,n+2,\dots\hspace{-.5pt},2n)\cdots\nn\\
&\hspace{1cm}\times(n^2-n+1,n^2-n+2,\dots\hspace{-.5pt},n^2)\,,\nn\vspace{2pt}\\\lb{perm}
&\pi_B :\; (1,n+1,\dots\hspace{-1pt},n^2-n+1)\\\nn
&\hspace{1cm}\times(2,n+2,\dots,n^2-n+2)\cdots(n,2n,\dots\hspace{-1pt},n^2)\,,\vspace{2pt}\\\nn
&\pi_C :\; (1)(2)(3)\cdots(n^2)\,.
\eea
Notice that $\pi_C$ is the identity. For $n=2$, the multientropy is related to other known information theoretic quantities such as the Rényi reflected entropy and CCNR \mbox{negativities,} see \cite{Dutta:2019gen,Penington:2022dhr,Milekhin:2022zsy,Berthiere:2023gkx}.

\pagebreak
We further introduce the genuine multientropy as the difference between the multientropy and the averaged sum of the Rényi entanglement entropies,
\bea\lb{def_genuine}
&G_n^{(3)}(A:B:C) = S^{(3)}_n(A:B:C) \\\nn
&\hspace{3cm} - \frac12\big(S_n(A)+S_n(B)+S_n(C)\big)\,,
\eea
where $S_n(K)=\frac{1}{1-n}\log\hspace{1pt}\Tr\rho_K^n$ is the Rényi entanglement entropy for subsystem $K$. 
This quantity is expected to characterize the amount of genuine tripartite entanglement (see, e.g., \cite{Penington:2022dhr,Harper:2024ker,Iizuka:2025ioc,Iizuka:2025caq}).
For $n=2$, the genuine multientropy reduces to $G_2^{(3)} = \frac12 M_{2,2}$, where $M_{m,n}=S^R_{m,n}-I_n$ is the generalized Markov gap \cite{Penington:2022dhr,Milekhin:2022zsy,Berthiere:2023gkx} given by the difference between the Rényi reflected entropy $S^R_{m,n}$ and the Rényi mutual information $I_n$. Several properties of (genuine) multientropy have been studied in, e.g., \cite{Gadde:2023zni,Gadde:2023zzj,Liu:2024ulq,Iizuka:2024pzm,Iizuka:2025bcc}.

Given such a family of entropies, a natural question is whether one can analytically continue to noninteger values of $n$---as argued in \cite{Gadde:2022cqi}---and obtain an entanglement measure in the limit $n\rightarrow 1$. 
For general states, there is no obvious approach to compute it---should it exists. Considering this difficulty, investigations have mostly been focusing on the $n=2$ Rényi multientropy, since it is easily computable in many contexts of interest.

In this paper, we present an analytical continuation of multientropy in quantum field theory. We consider Lifshitz theory which provides a formidable playground due to its Rokhsar-Kivelson (RK) groundstate.

\medskip
\textit{Dihedral invariant} --- 
Dihedral invariants \cite{Gadde:2024taa,Harper:2025uui} are a family of local unitary invariants on three parties introduced as new potential probes for investigating multipartite entanglement in quantum many-body systems (see also \cite{Gadde:2025csh}). They are formed by taking $2n$ copies of the density matrix and performing contractions of the three parties with respect to a set of permutations associated with the dihedral group $\mathbb{D}_{2n}$
\begin{equation}
    \mathbb{D}_{2n} : \langle\spp r,s\spp|\spp r^n\spm\spm=\spm s^2\spm\spm=\spm e\spp; sr\spm=\spm r^{n-1}s\spp \rangle\spp,
    \label{DihedralGroup}
\end{equation}
where $e$ the identity element of $\mathbb{S}_{2n}$.

The partition function associated with the dihedral-symmetric $2n$ replicas is expressed as 
\bea
\mathcal{Z}_{2n}=\langle\Psi|^{\otimes 2n}\big(a^D\otimes b^D \otimes c^D\big)|\Psi \rangle^{\otimes2n}\spp,
\eea
where $a^D, b^D, c^D$ are in $\mathbb{D}_{2n}$ and permute the $2n$ copies of subsystems $A,B,C$, respectively. Their explicit form read \cite{Harper:2025uui}
\begin{equation}
\begin{aligned}
    &a^D: \; (1,2,...,n)(n+1,n+2,...,2n)\,,\\
    &b^D: \; (1,2n)(2,2n-1)(3,2n-2)\cdots(n,n+1)\,,\\
    &c^D: \; (1)(2)(3)\cdots (2n)=e\,.
\end{aligned}
\end{equation}
From the above partition function with dihedral symmetry, one may construct the following \mbox{dihedral} invariant 
\bea
\begin{split}
\mathcal{D}_{2n}(A:B)= \frac{1}{1-n}\log \frac{\mathcal{Z}_{2n}}{(\mathcal{Z}_2)^n}\,.
\end{split}
\lb{di}
\eea
The partition function is normalized such that $\lim_{n\to1}\mathcal{Z}_{2n}/(\mathcal{Z}_2)^n=1$. Different normalizations are possible (see, for example, \cite{Gadde:2024taa,Harper:2025uui}, and Appendix \ref{Apdx_norm_di} for a definition involving another normalization), but this particular choice will later reveal to be natural. Note that, besides normalization, \eqref{di} also differs from the definition in \cite{Harper:2025uui} by an unimportant factor $1/2n$. Here, we study the dihedral invariant \eqref{di} for general pure tripartite states, and illustrate our universal results with Lifshitz groundstates, GHZ and W states.

\begin{figure}[t]
\centering
\includegraphics[scale=0.94]{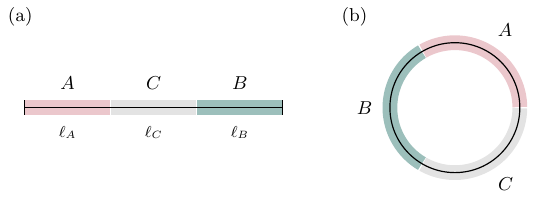}
\vspace{-15pt}
\caption{Examples of tripartitions, with subsystem sizes $\ell_K$. (a) A system with boundary with $A, B$ disjoint. (b) A periodic system with contiguous $A, B, C$.}
\lb{fig_tri}
\end{figure}

\begin{figure*}[t]
\centering
\includegraphics[scale=1.25]{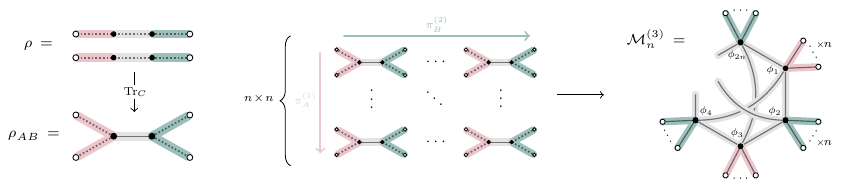}
\vspace{-5pt}
\caption{Graphical representation of the calculations leading to the multientropy for $A, B$ disconnected in a finite interval (see Fig.\,\hyperref[fig_tri]{\ref{fig_tri}(a)}). We use different colors to differentiate lengths $\ell_A$ (pink), $\ell_B$ (green), and $\ell_C$ (gray). Dotted lines denote the ``indices'' of the matrix, while solid lines mean that trace has been taken. Each field value at the interface between the subsystems (black dots), denoted $\phi_i$, has to be integrated over. Hollow circles indicate Dirichlet boundary conditions at the end of the system.}
\lb{fig_GDD}
\end{figure*}

\medskip
\textit{Lifshitz groundstates} --- 
In this work, we focus on a particular class of nonrelativistic quantum field theories known as Lifshitz theories, distinguished by the remarkable property that their groundstate wavefunctional assumes a local form, given in terms of the action of a classical model. Lifshitz groundstates are representatives of Rokhsar-Kivelson states \cite{Rokhsar:1988zz,henley2004classical} which quantum mechanically encode the partition function of a classical system. 

We shall consider the $z=2$ Lifshitz critical boson and its massive deformation in $1+1$ dimensions \cite{Boudreault:2021pgj}, described by the Hamiltonian 
\bea\label{E:H_Lifshitz}
H=\frac{1}{2}\int d x \left(\Pi^2 + (\nabla^2 \phi)^2 + 2 \omega^2 (\nabla \phi)^2 + \omega^4 \phi^2 \right),
\eea
with canonical commutation relations $[\phi(x),\Pi(x')]=i\delta(x-x')$, and $\Pi(x)=-i\delta / \delta\phi(x)$ in the Schrödinger picture.
The groundstate can be expressed in terms of a path integral of a one-dimensional Euclidean theory
\bea\label{E:Psi}
\begin{split}
&\ket{\Psi}=\frac{1}{\sqrt{\mathcal{Z}}}\int \D \phi \,e^{-\frac{1}{2}S_{\rm cl}[\phi]}\ket{\phi},\\
&\,S_{\rm cl}[\phi]=\int dx \big((\nabla\phi)^2+\omega^2\phi^2\big)\spp,
\end{split}
\eea
with normalization factor $\mathcal{Z}=\int \mathcal{D}\phi\, e^{-S_{\rm cl}[\phi]}$.
In one dimension, $\mathcal{Z}$ corresponds to the partition function of a single particle with classical action $S_{\rm cl}[\phi]$, i.e.~an Euclidean harmonic oscillator of ``mass" $M=2$ and ``frequency" $\omega$. In the critical massless case, $\omega=0$, the theory is invariant under Lifshitz scaling with $z=2$, and the classical action is that of a free nonrelativistic particle.

\pagebreak
The propagator of the Euclidean harmonic oscillator, defined as $K(\phi, \phi';\ell)=\int_{\phi(0)=\phi}^{\phi(\ell)=\phi'}\D\phi\, e^{-S_{\rm cl}[\phi]}$, reads
\bea\label{E:propagator_mass}
\begin{aligned}
&K(\phi, \phi';\ell)= \\
&\qquad\sqrt{\frac{\omega}{\pi  \sinh \omega \ell}}\exp\bigg[
					\frac{-\omega\big((\phi^2 + {\phi'}^2)\cosh \omega\ell - 2\phi\phi'\big)}{\sinh\omega \ell}\bigg],
\end{aligned}
\eea 
and reduces to that of the free nonrelativistic particle in the massless limit $\omega=0$.
Vacuum expectation values associated to $\ket{\Psi}$ of local operators, such as partition functions, can then be expressed in terms of the above propagator. 
Various entanglement properties of Lifshitz theories have been studied in, e.g., \cite{Fradkin:2006mb,Hsu:2008af,PhysRevB.80.184421,Oshikawa:2010kv,2011PhRvL.107b0402Z,Zhou:2016ykv,Chen:2016kjp,chen2017quantum,MohammadiMozaffar:2017nri,MohammadiMozaffar:2017chk,Angel-Ramelli:2019nji,Angel-Ramelli:2020wfo,Angel-Ramelli:2020xvd,Boudreault:2021pgj,Parez:2022ind}.

\subsection{Summary of main results}

\textit{Genuine multientropy} --- 
We compute the multientropy for tripartite Lifshitz groundstates. We consider adjacent and disconnected subsystems $A, B$, and different boundary conditions.  In all cases investigated, we find that the genuine multientropy obeys the relation
\bea\lb{GMgen_intro}
G_n^{(3)}(A:B:C) = \frac{2-n}{2n}\Big(I_{1/2}(A:B)-2\E(A:B)\Big),
\eea
where $I_{1/2}$ is the $n=1/2$ Rényi mutual information, and $\E=\log||\rho_{AB}^{T_B}||$ is the logarithmic negativity \cite{Vidal:2002zz}. For $n=2$, the genuine multientropy vanishes. This is in agreement with the vanishing of the generalized Markov gap $M_{2,2}$ for Lifshitz groundstates since $G_2^{(3)} = \frac12 M_{2,2}$, see \cite{Berthiere:2023bwn}. Note that for disjoint $A, B$, the reduced state is separable and hence the logarithmic negativity is zero \cite{Boudreault:2021pgj,Angel-Ramelli:2020wfo}. The genuine multientropy \eqref{GMgen} is nonnegative for $n\le2$ and nonpositive for $n>2$, and it is UV--finite.

\medskip
\textit{Dihedral invariant} --- 
For general tripartite pure states, we demonstrate that dihedral invariants \eqref{di} are exactly Rényi reflected entropies, i.e.
\begin{equation}
    \mathcal{D}_{2n}(A:B)=S_{2,n}^R(A:C)\spp.
\end{equation}
In particular, we show that the dihedral permutations of replicas are equivalent to the reflected construction, or alternatively to the realignment of density matrices. For $n=2$, dihedral invariant and multientropy are then simply related as $S^{(3)}_2(A:B:C)=\frac12\D_4(A:B)+S_2(B)$.

\section{Genuine multientropy\\for Lifshitz groundstates}\lb{sec:GM}

We compute the multientropy using the approach \mbox{developed} in \cite{Boudreault:2021pgj,Berthiere:2023bwn} for the $(1+1)$--dimensional $z=2$ Lifshitz field theory. This method yields the partition function $\Z_n^{(3)}$ on a replica graph $\M_n^{(3)}$ in terms of the propagator of the harmonic oscillator \eqref{E:propagator_mass}, which can be evaluated for any integer $n$ and, as we shall show, analytically continued to noninteger values.

The basic ingredient is the reduced groundstate density matrix on $A\cup B$, of which we take $n^2$ copies. Replicas are then glued together according to the permutations rule in \eqref{perm}. To better visualize how this is done, one may organize the $n^2$ replicas as an $n\times n$ grid, numbered vertically from top to bottom and from left to right. That way, traces over $A$ are taken vertically on each column, while traces over $B$ are taken horizontally on each row. We use dotted lines for bras and kets, and solid lines to represent traces taken, which simply are propagators (see Fig.\,\ref{fig_GDD} for an example). Black dots correspond to field degrees of freedom that must be integrated over. The resulting replica graph depends on the tripartition of the system. In general, the partition function on the replica graph is a Gaussian matrix integral,
\bea
\begin{split}
\Z_n^{(3)} &= \frac{1}{\big(\prod_{K}\frac{\pi}{\o}\sinh(\omega \ell_K)\big)^{n^2/2}} \int d\boldsymbol{\phi}\,e^{-\boldsymbol{\phi}^T M_n^{(3)}\boldsymbol{\phi}}\\
&=\frac{1}{\big(\prod_{K}\frac{\pi}{\o}\sinh(\omega \ell_K)\big)^{n^2/2}} 
\sqrt{\frac{1}{\det(M_n^{(3)}/\pi)}}\,,
\end{split}
\eea
where $K\in\{A,B,\dots\}$, and the form of the matrix $M_n^{(3)}$ stems from the replica graph $\M_n^{(3)}$.

Below, we illustrate the replica method for the multientropy with two tripartitions of a one-dimensional system. More general cases can be found in Appendix \ref{Apdx:GME}.

\subsection{Disjoint subsystems $A, B$ on an interval}

Let us first consider the configuration shown in Fig.\,\hyperref[fig_tri]{\ref{fig_tri}(a)}, where $A$ and $B$ are adjacent to the boundaries of the system. We work on a finite one-dimensional system with Dirichlet boundary conditions at both ends. The partition function $\Z_{n}^{(3)}$ of the harmonic oscillator on the replica graph $\M_{n}^{(3)}$ can be calculated using the procedure shown in Fig.\,\ref{fig_GDD}, yielding
\bea
&\mathcal{Z}_n^{(3)}=\int d\phi_1 ...\hspace{1pt} d\phi_{2n} \, \prod_{i=1}^{n}K(0,\phi_{2i};\ell_A)^n K(0,\phi_{2i-1};\ell_B)^n  \nn \\[-0.18cm]
&\hspace{1.255cm}\times\spm\spm\prod_{j=1}^{2n-1}\hspace{-2pt}K(\phi_j,\phi_{j+1};\ell_{C})\hspace{-2pt}\prod_{k=1}^{\hspace{2pt}\big\lceil\spm \spm \frac{2n-j}{2} \spm\spm\big\rceil \spm-1}\hspace{-7pt}K(\phi_j,\phi_{j+2k+1};\ell_{C})  \nn\\
&\hspace{0.65cm}=\frac{(\o/\pi)^{3n^2/2}\sqrt{\pi^{2n}/\det M_n^{(3)}}}{\big(\spm\sinh(\omega \ell_A)\sinh(\omega \ell_B)\sinh(\omega \ell_C)\big)^{n^2/2}}\,,
\eea
where $M_{n}^{(3)}$ is a $2n\times 2n$ matrix given in (\ref{MatDis}). Its determinant can be evaluated for any integer $n$,
\bea
\det M_n^{(3)}\hspace{-1.5pt}=\hspace{-.5pt} (n\omega)^{2n}\frac{\sinh(\omega \ell) \big(\hspace{-1pt}\sinh(\omega \ell_{AC})\sinh(\omega \ell_{BC})\big)^{n-1}}{\sinh(\omega \ell_A)^n\sinh(\omega \ell_B)^n\sinh(\omega\ell_C)^{2n-1}},
\eea
and provides an analytic continuation to nonintegers $n$. Here $\ell=\ell_A+\ell_B+\ell_C$, $\ell_{AB}=\ell_A+\ell_B$, and so on.
The Rényi multientropy easily follows from \eqref{def_multi}.

Since we are mostly interested in the genuine multientropy \eqref{def_genuine}, let us directly present its expression
\bea\lb{GM_dis_DD1}
G_n^{(3)}(A:B:C)= \frac{2-n}{4n}\log \frac{\sinh(\omega \ell_{AC}) \sinh(\omega \ell_{BC})}{\sinh(\omega\ell_C)\sinh(\omega \ell)}\,.
\eea
The bipartite Rényi entropies used to calculate it can be found in \cite{Boudreault:2021pgj}.
We observe the genuine multientropy \eqref{GM_dis_DD1} is nonnegative for $n\le2$ and nonpositive for $n>2$, and it is UV--finite. Further, the dependence on the Rényi index $n$ completely factorizes and such that the genuine multientropy vanishes for $n=2$. This is in agreement with the vanishing of the generalized Markov gap $M_{2,2}$ since we have the relation $G_2^{(3)} = \frac12 M_{2,2}$, see \cite{Berthiere:2023bwn}. 

\medskip
\textit{Taking limits} --- In the massless limit, we find 
\bea
G_n^{(3)}= \frac{2-n}{4n}\log \frac{\ell_{AC} \ell_{BC}}{\ell_C\ell}\,,\qquad \o\rightarrow0\,.
\eea
In the highly gapped regime $\omega \ell_i\rightarrow \infty$ for all $\ell_i$, the genuine multientropy vanishes exponentially
\bea
G_n^{(3)}\simeq \frac{2-n}{4n} e^{-2\o\ell_C}\,, \qquad \o\ell_i\rightarrow\infty\,.
\eea
Note that in both regimes, $G_n^{(3)}$ goes to zero as $\ell_C\rightarrow\infty$, and vanishes for $A$ or $B\rightarrow\emptyset$. This is coherent with the expectation that genuine multientropy characterizes the amount of tripartite entanglement, since in the first case $A$ and $B$ become infinitely separated effectively yielding a biseparable state, while in the second case we have a bipartite state. In contrast, $G_n^{(3)}$ diverges as $\ell_C$ goes to zero, similarly to mutual information.

\subsection{Adjacent subsystems $A, B$ on a circle}

We now consider $A$ and $B$ adjacent on a circle, as in Fig.\,\hyperref[fig_tri]{\ref{fig_tri}(b)}.
The replica manifold is similar as that in Fig.\,\ref{fig_GDD}, only the boundary vertices (hollow dots), which represent Dirichlet boundary conditions, must pair together.
We draw the replica graph $\M_{3}^{(3)}$ in Fig.\,\ref{AdjPeriodic}, which is straightforwardly generalized to arbitrary $n$. The matrix $M_{n}^{(3)}$ entering the corresponding partition function is given in (\ref{MatAdj2}). Its determinant can be evaluated for any $n$,
\bea
&\det M_n^{(3)}=4(n\omega)^{3n}\sinh(\omega \ell/2)^2 \\\nn
&\qquad\qquad\;\times\spm\spm\frac{\big(\spm\sinh(\omega \ell_{AB})\sinh(\omega \ell_{BC})\sinh(\omega \ell_{AC})\big)^{n-1}}{\big(\spm\sinh(\omega \ell_A)\sinh(\omega \ell_B)\sinh(\omega \ell_{C})\big)^{2n-1}}\spm.
\eea
The genuine multientropy for adjacent subsystems on a circle hence follows as
\bea\lb{GM_adj_P1}
&G_n^{(3)}=\\\nn
&\frac{2-n}{4n}\log \frac{\sinh(\omega \ell_{AB})\sinh(\omega \ell_{BC})\sinh(\omega \ell_{AC})}{4\sinh(\omega\ell_{A})\sinh(\omega\ell_{B})\sinh(\omega\ell_{C})\sinh(\omega \ell/2)^2}\,.
\eea

\medskip
\textit{Taking limits} --- In the massless limit, we find 
\bea
G_n^{(3)}= \frac{2-n}{4n}\Big(\spm\log \frac{\ell_{AB} \ell_{BC} \ell_{AC}}{ \ell_{A} \ell_{B}\ell_C}-2\log\o\ell\Big)\,,\quad \o\rightarrow0\,.
\eea
We observe a logarithmic divergence as $\o\rightarrow0$ due to the zero mode of the massless theory on a circle.  
In the highly gapped regime $\omega \ell_i\rightarrow \infty$ for all $\ell_i$, the genuine multientropy vanishes exponentially.
Note that in both regimes, $G_n^{(3)}$ diverges as any $\ell_i$ goes to zero.

\subsection{A general formula for Lifshitz groundstates}

In all investigated cases (see Appendix \ref{Apdx:GME} for other configurations), the genuine multientropy can be written in terms of partition functions of the Euclidean (massive) scalar field, i.e.
\bea
G_n^{(3)}(A:B:C) = -\frac{2-n}{2n}\log\frac{\Z_{AB}\Z_{AC}\Z_{BC}}{\Z_A\Z_B\Z_C\Z},
\eea
where $\Z_K$ is the partition function on region $K$ with Dirichlet boundary conditions at the interface with its complement. We further find that the genuine multientropy obeys the following relation
\bea\lb{GMgen}
G_n^{(3)}(A:B:C) =\frac{2-n}{2n}\Big(I_{1/2}(A:B)-2\E(A:B)\Big) ,
\eea
where $I_{n}(A:B)= S_n(A) + S_n(B) - S_n(A\cup B)$ is the Rényi mutual information, and $\E(A:B)=\log||\rho_{AB}^{T_B}||$ is the logarithmic negativity \cite{Vidal:2002zz} with $\rho_{\spm AB}^{T_B}$ the partial transpose \cite{Peres:1996dw} with respect to $B$.
For Lifshitz groundstates, the mutual information reads \cite{Fradkin:2006mb,Boudreault:2021pgj}
\bea
I_{n}(A:B)=-\log\frac{\Z_A\Z_B\Z_{AC}\Z_{BC}}{\Z_{AB}\Z_C\Z}+ 2c(n)\,,
\eea
and the logarithmic negativity \cite{Angel-Ramelli:2020wfo}
\bea
\E(A:B)=-\log\frac{\Z_A\Z_B}{\Z_{AB}}+ c(1/2)\,,
\eea
where $c(n)=(p_A+p_B-p_{A\cup B})\frac{\log n}{4(n-1)}$ with $p_K$ being the number of boundaries separating $K$ from the rest of the system. It is then straightforward to check relation \eqref{GMgen} for all considered cases (see also Appendix \ref{Apdx:GME}). We conjecture \eqref{GMgen} to hold for general tripartition of Lifshitz groundstates. Whether relation \eqref{GMgen} holds in other contexts is an intriguing question. 


\subsection{Comparison with the Markov gap}

Genuine multientropy was introduced with the aim of quantifying multipartite entanglement. In particular, for tripartite pure states, it is expected to detect both GHZ and W entanglement (see, e.g., \cite{Gadde:2022cqi}). In contrast, the Markov gap \cite{Hayden:2021gno} (the difference between reflected entropy and mutual information) whose positivity signals genuine tripartite entanglement \cite{Zou:2020bly}, is blind to GHZ entanglement. As discussed in, e.g., \cite{Liu:2024ulq,Iizuka:2025ioc}, genuine multientropy and Markov gap share similar features. It is therefore interesting to compare the two in the context of continuum RK states. 

To do so, let us take a look at the tripartition shown in Fig.\,\hyperref[fig_tri]{\ref{fig_tri}(a)} with $A, B$ disjoint, for which we have ($n\rightarrow1)$
\bea\lb{GM_adj_Pn1}
&G^{(3)}=\frac{1}{4}\log\frac{\sinh(\omega \ell_{AC}) \sinh(\omega \ell_{BC})}{\sinh(\omega\ell_C)\sinh(\omega \ell)}\,,\\
&M=\frac{1}{\sqrt{1-\eta}}\log\spm\(\frac{1+\sqrt{1-\eta}}{\sqrt\eta}\)  +\frac12\log(\eta/4)\,,
\eea
where $\eta=\frac{\sinh\o\ell_{A}\sinh\o\ell_{B}}{\sinh\o\ell_{AC}\sinh\o\ell
_{BC}}$ \cite{Berthiere:2023bwn}. 
Both quantities vanish when sending either $\ell_A$ or $\ell_B$ to zero. However, in the limit of small separation $\ell_C\rightarrow0$, $G^{(3)}$ diverges while $M$ remains finite.
On a periodic system, in the critical massless limit $G^{(3)}$ diverges logarithmically with $\o$. This is due to the zero mode of the massless theory on a circle. In contrast, the Markov gap remains finite. In general, the Markov gap does not suffer from any divergence for Lifshitz groundstates \cite{Berthiere:2023bwn}.

\section{Dihedral Invariant}\lb{sec:dihedral}

We show that for tripartite pure states, the dihedral invariant \eqref{di} is none other than the \mbox{$(2,n)$-Rényi} reflected entropy. The reflected entropy was introduced in \cite{Dutta:2019gen} and further studied in \cite{Jeong:2019xdr,Bueno:2020vnx,Li:2020ceg,Berthiere:2020ihq,Bueno:2020fle,Akers:2022max,Chen:2022fte,Vasli:2022kfu,Lu:2022cgq,Afrasiar:2022fid,Hayden:2023yij,Sohal:2023hst,Basak:2023uix,Berthiere:2023bwn}. This quantity has a meaningful relationship to entanglement, in particular with tripartite entanglement \cite{Akers:2019gcv,Hayden:2021gno,Zou:2020bly}, and its $(m,n)$--Rényi generalization is related to other known quantities as discussed in \cite{Berthiere:2023gkx}.
Let us begin by presenting their definitions in terms of permutation operators.

\subsection{Definitions in terms of permutation operators}
First, we repeat that of the dihedral invariant \eqref{di} for the reader's convenience
\begin{equation}\lb{di-inv}
\begin{aligned}
    &\mathcal{D}_{2n}(A:B)= \frac{1}{1-n}\log \frac{\mathcal{Z}_{2n}}{(\mathcal{Z}_2)^n}\,, \\[1ex]
    &\mathcal{Z}_{2n}=\langle\Psi|^{\otimes 2n}\big(a^D\otimes b^D \otimes e\big)|\Psi \rangle^{\otimes2n}\,,
\end{aligned}
\end{equation}
with permutation operators
\begin{equation}
\begin{aligned}
    &a^D:\spm\spm\spm & &(1,2,...,n)(n+1,n+2,...,2n)\,,\\[.25ex]
    &b^D:\spm\spm\spm & &(1,2n)(2,2n-1)(3,2n-2)\cdots(n,n+1)\,,
\end{aligned}
\end{equation}
of the replicas of subsystems $A$ and $B$, respectively, and $e$ the identity on subsystem $C$.
Second, the $m=2$ reflected entropy may be defined in terms of permutation operators as \cite{Dutta:2019gen,Liu:2024ulq}
\begin{equation}\lb{SRdef}
\begin{aligned}
    &S^R_{m=2,n}(A:C)= \frac{1}{1-n}\log \frac{\mathcal{Z}_{2,n}}{(\mathcal{Z}_{2,1})^n}\,, \\[1ex]
    &\mathcal{Z}_{2,n}=\langle\Psi|^{\otimes 2n}\big(a^R\otimes e\otimes c^R\big)|\Psi \rangle^{\otimes2n}\,,
\end{aligned}
\end{equation}
with 
\begin{equation}
\begin{aligned}
    &a^R:\spm\spm\spm & &(1,2)(3,4)\cdots(2n-1,2n)\,,\\[.25ex]
    &c^R:\spm\spm\spm & &(2,3)(4,5)\cdots(2n,1)\,.
\end{aligned}
\end{equation}
Both quantities may thus be defined from their permutation elements acting on $A$, $B$ and $C$ as follows
\begin{equation}
\begin{aligned}
    &\mathcal{D}_{2n}(A:B)\hspace{-8pt} &: \;\; &(a^D,b^D,e)\spp, \\[.25ex]
    &S^R_{2,n}(A:C)\hspace{-8pt} &: \;\; &(a^R,e,c^R)\spp.
\end{aligned}
\end{equation}

\subsection{Proof of $\D_{2n}(A:B)=S^R_{2,n}(A:C)$}

Freedom of relabeling either the kets, bras or both simultaneously grants the equivalence relations \cite{Gadde:2022cqi,Gadde:2024taa,Gadde:2023zzj}
\begin{equation}
\begin{aligned}
        &(\sigma_1,\sigma_2,\sigma_3) \sim  (\sigma_1,\sigma_2,\sigma_3)\, g\,,\quad &g\in \mathbb{S}_{2n} \spp ,\\[.25ex]
    &(\sigma_1,\sigma_2,\sigma_3) \sim  h^{-1}(\sigma_1,\sigma_2,\sigma_3)\,h\,,\;\; &h\in \mathbb{S}_{2n} \spp.
\end{aligned}
\end{equation}
Acting with $c^R$ on the right in $\Z_{2,n}$ in \eqref{SRdef} such as to recover a trace on $C$ yields
\begin{equation}
    \mathcal{S}^R_{2,n}(A:C) : \; (a^R,e,c^R) \sim (a^Rc^R,c^R,e) \,,
    \label{DonC2}
\end{equation}
and defining the new set of permutation elements 
\begin{equation}
    (a^R c^R\spm\equiv t, c^R, e)\spp,
\end{equation}
one can build the symmetry group
\begin{equation}
    G:\;\langle\spp t,c^R\spp|\spp t^n\spm\spm=\spm (c^R)^2\spm\spm=\spm e\spp; c^R t\spm=\spm t^{n-1}c^R\spp \rangle\spp.
\end{equation}
Since $G$ has order $2n$, there exists an isomorphism \mbox{$f\;:\;\mathbb{D}_{2n}\spm\spm\rightarrow G$.} Note that it is equivalent to act on the left with $c^R$, and the above argument carries through.

We look for the explicit form of the isomorphism $f$ as a conjugation to preserve the identity trace on $C$. This is equivalent to finding an $h\in \mathbb{S}_{2n}$ such that
\begin{equation}
    h^{-1}(a^D,b^D,e)\,h=(t,c^R,e) \,.
    \lb{EqRel2}
\end{equation}
It can indeed be shown that
\begin{equation}
\begin{aligned}
&h: \; \left(
\begin{matrix}
1 & 2 & 3 & \cdots & 2n\\
h(1) & h(2) & h(3) & \cdots & h(2n)
\end{matrix} \right)\spm,\\[.75ex]
    &h(j) = \left \{ \begin{matrix}
        1, && j=n\spp, \\[.35ex]
        2j+1, && 1\leq j\leq n-1\spp, \\[.35ex]
        4n+2-2j\spp, && n+1\leq j \leq 2n\spp,
    \end{matrix} \right .
\end{aligned}
\end{equation}
satisfies \eqref{EqRel2}. As before, it is equivalent to act on the left with $c^R$ in \eqref{DonC2}. Doing so yields the set of equations
\bea
     (h')^{-1} a^D \, h' =c^Ra^R\,, \qquad (h')^{-1} b^D\, h' =c^R\,,
\eea
solved by $h'=h\,c^R$. 
This completes our proof that the replica groups associated with \mbox{$\mathcal{D}_{2n}\spm(A\spm:\spm B)$ and $S^R_{2,n}\spm(A\spm:\spm C)$} are isomorphic, hence
\begin{equation}\lb{rel_di_SR}
    \mathcal{D}_{2n}(A:B)=S_{2,n}^R(A:C)\spp,
\end{equation}
with the given choice of normalization. \hfill $\square$

The unnormalized dihedral partition function $\log\Z_{2n}$ can alternatively be expressed in terms of the realignment of reduced density matrices \cite{2002quant.ph..5017C}, and can be identified with the Rényi CCNR negativity 
\bea\lb{di-ccnr}
\log\Z_{2n} = \E_n^{\scalebox{0.6}{\rm CCNR}}(A:C) \spp.
\eea
Indeed, the realignment of density matrices is equivalent to the permutations for reflected entropy \cite{Berthiere:2023gkx,Yin:2022toc,Milekhin:2022zsy}. The above quantity is based on the CCNR criterion \cite{2003JPhA...36.5825R,2002quant.ph..5017C} that states $\E^{\scalebox{0.6}{\rm CCNR}}_{1/2} \leqslant 0$ for separable states.

As a consequence of relation \eqref{rel_di_SR}, the family of dihedral invariants \eqref{di} for $n\in (0,2)$ is not monotonically nonincreasing under partial trace, as was proven \cite{Berthiere:2023gkx} for the Rényi reflected entropy in the range $m>0$ and $n\in (0,2)$. Hence, it does not qualify as a correlation measure.

In Appendix \ref{dihedral}, we provide checks of relation \eqref{rel_di_SR} for Lifshitz groundstates, GHZ and W states.
The CFT \mbox{calculation} of \cite{Harper:2025uui}, up to normalization, is in agreement with \eqref{rel_di_SR}, and the excess the authors define there is \mbox{actually} proportional to the generalized Markov gap.


\vspace{20pt}
\section{Discussion}\lb{sec:conclu}

We have investigated the (genuine) multientropy in Lifshitz theories. This quantity is notoriously difficult to compute, even for pure three-qubit states, and there is no obvious analytical continuation to noninteger Rényi index in general. The literature thus mostly focuses on the $n=2$ Rényi multientropy since it is easily calculable. In this work, we succeed in computing the multientropy for arbitrary Rényi index in Lifshitz theories.  We have found that the (genuine) multientropy is nonnegative for $n\le2$ and UV--finite. For general tripartite pure states, the (genuine) multientropy is expected \cite{Penington:2022dhr,Iizuka:2025ioc} -- but not proven -- to be nonnegative for $n<2$, our results thus provide further evidence in that direction.

Our main result is formula \eqref{GMgen} which relates the genuine multientropy for Lifshitz groundstates to two well-known bipartite quantities, namely the $1/2$-Rényi mutual information and the logarithmic negativity. Interestingly, relation \eqref{GMgen} also holds for tripartite states with separable bipartite marginal of the form $\rho_{AB}=\frac1d\sum_{i=1}^{d} \rho_A^i\otimes\rho_B^i$, where the $\rho_K^i$ are proportional to projectors with orthogonal support. Such states include GHZ states. Furthermore, relation \eqref{GMgen} is satisfied for tripartite pure stabilizer states, which are local-unitary equivalent \cite{Bravyi:2005ztn,Looi:2011jrm,Wong:2025cwc} to a collection of GHZ states, Bell pairs, and unentangled single-qudits. Indeed, the difference $I_{1/2}-2\E$ counts the number of GHZ-states that can be extracted from a stabilizer state, and so does the genuine multientropy (see, e.g., \cite{Akella:2025owv,Yuan:2025dgx} for recent related studies). However, the nature of the tripartite entanglement in Lifshitz groundstates is not purely GHZ-like since the Markov gap is positive \cite{Berthiere:2023gkx} (it vanishes for GHZ states \cite{Zou:2020bly}). Though relation \eqref{GMgen} does not hold for general states, whether it does in different contexts, and why that is, are interesting questions worth investigating.

We have studied another class of multi-invariants, with dihedral replica symmetry, for general quantum states on three parties $A, B, C$. We have shown that the dihedral invariant defined in \eqref{di} is exactly related to the Rényi reflected entropy. In particular, we have demonstrated that the symmetry group of the replicas for reflected entropy is isomorphic to the dihedral group such that the ``dihedral permutations'' of subsystems $A$ and $B$ are equivalent to the ``reflected permutations'' of $A$ and $C$, yielding relation \eqref{rel_di_SR}. It would be interesting to explore further the landscape of multi-invariants and their relationships.

\section*{Acknowledgments}
It is a pleasure to thank Jonathan Harper for interesting discussions regarding multi-invariants, as well as Hongjie Chen, Gilles Parez, Shinsei Ryu and Yang Zhou.

\onecolumngrid
\clearpage
\appendix

\setcounter{figure}{0}
\makeatletter
\renewcommand{\thefigure}{\thesection\arabic{figure}}

\section{Genuine multientropy: more tripartition configurations}
\lb{Apdx:GME}

\subsection{On a finite interval}

\subsubsection{Disjoint $A, B$ in the bulk}

\begin{figure*}[ht]
\centering
\includegraphics[scale=1]{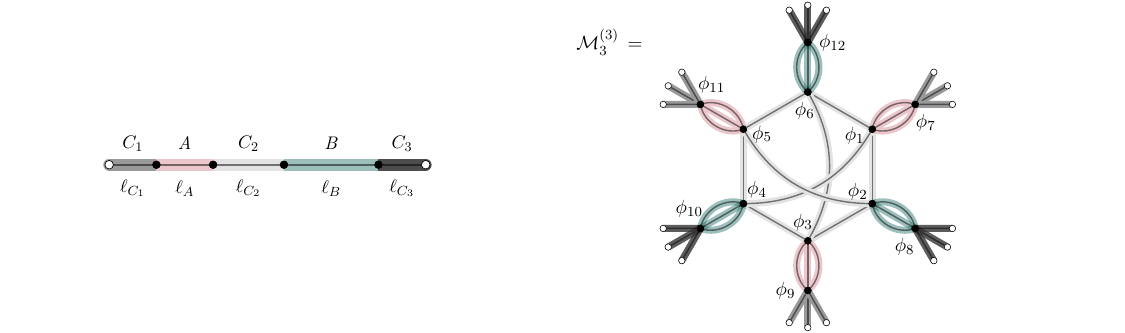}
\caption{Disjoint regions $A$ and $B$ in the bulk. Right: Replica graph resulting from the multientropy symmetry for $n=3$.}
\lb{DisBulk}
\end{figure*}

\textit{Multientropy} ---
We begin by computing the multientropy for Dirichlet boundary conditions, and then briefly cover Neumann boundary conditions. For this calculation only, we reinstate the UV regulator $\epsilon$, and will see that it cancels out in the genuine multientropy. We draw the replica graph $\M_{3}^{(3)}$ in Fig.~\ref{DisBulk}, which is straightforwardly generalized to arbitrary $n$ and yields
\begin{equation}
\begin{aligned}
\mathcal{Z}_n^{(3)} &= \int d\phi_1 ... d\phi_{4n} \prod_{i=1}^{n} \Big(K(0,\phi_{2n+2i-1};\ell_{C_1})K(\phi_{2n+2i-1},\phi_{2i-1};\ell_A)\Big)^n \Big (K(0,\phi_{2n+2i};\ell_{C_3})K(\phi_{2n+2i},\phi_{2i};\ell_B) \Big)^n \\
& \hspace{2.5cm}\times\prod_{j=1}^{2n-1} \hspace{-2pt}K(\phi_j,\phi_{j+1};\ell_{C_2})\hspace{-2pt}\prod_{k=1}^{\hspace{2pt}\big\lceil\spm \spm \frac{2n-j}{2} \spm\spm\big\rceil \spm-1}\hspace{-7pt}K(\phi_j,\phi_{j+2k+1};\ell_{C_2}) \\
&=  \Bigg(\frac{(\omega\epsilon/\pi)^5}{\sinh(\omega \ell_A) \sinh(\omega \ell_B) \sinh(\omega \ell_{C_1})\sinh(\omega \ell_{C_2})\sinh(\omega \ell_{C_3})} \Bigg)^{n^2/2} \sqrt{\frac{\pi^{4n}}{\det M_{n}^{(3)}}}\spp,
\end{aligned}
\end{equation}
where $M_{n}^{(3)}$ is a $4n \times 4n$ matrix given in (\ref{MatDis2}), for which the determinant is evaluated for any $n\in \mathbb{N}$ to
\begin{equation}
    \det M_{n}^{(3)}=(n\omega \epsilon)^{4n}\frac{\sinh(\omega \ell) \big(\spm\sinh(\omega \ell_{AC_1C_2}) \sinh(\omega \ell_{BC_2C_3})\big)^{n-1}}{\big(\spm\sinh(\omega \ell_A)\sinh(\omega \ell_B)\sinh(\omega \ell_{C_1})\sinh(\omega \ell_{C_3})\big)^n \sinh(\omega \ell_{C_2})^{2n-1}}.
\end{equation}
The resulting multientropy is
\begin{equation}
\begin{aligned}
    S^{(3)}_n(A:B:C)&=\log \frac{\sinh(\omega \ell_A)^{\frac{1}{2}}\sinh(\omega \ell_B)^{\frac{1}{2}}\sinh(\omega \ell_{C_1})^{\frac{1}{2}}\sinh(\omega \ell_{C_3})^{\frac{1}{2}}\sinh(\omega \ell_{AC_1C_2})^{\frac{1}{2n}}\sinh(\omega \ell_{BC_2C_3})^{\frac{1}{2n}}}{\omega^2 \epsilon^2 \sinh(\omega \ell_{C_2})^{\frac{1-n}{2n}}\sinh(\omega \ell)^{\frac{n+1}{2n}}}\\
& \hspace{1.5cm} + 2\log \pi + 2\frac{\log n}{n-1} \spp.
\end{aligned}
\end{equation}

\medskip
\textit{Genuine multientropy} --- The genuine multientropy is obtained by subtracting the averaged sum of the Rényi entanglement entropies from the multientropy. We have
\begin{equation}
    \begin{aligned}
&\frac{1}{2}\big(S_n(A)+S_n(B)+S_n(C) \big)=\\\nn
&\qquad \quad \frac{1}{4} \log \frac{\sinh(\omega \ell_{C_1AC_2})\sinh(\omega \ell_{C_2BC_3})\sinh^2(\omega \ell_{C_1})\sinh^2(\omega \ell_A)\sinh(\omega \ell_{C_2})\sinh^2(\omega \ell_{B})\sinh^2(\omega \ell_{C_3})}{\omega^8\epsilon^8\sinh^3(\omega \ell)}+ 2\log \pi + 2\frac{\log n}{n-1} \spp,
\end{aligned}
\end{equation}
and the genuine multientropy follows as
\begin{equation}
G_n^{(3)}(A:B:C)=\frac{2-n}{4n}\log \frac{\sinh(\omega \ell_{AC_1C_2})\sinh(\omega \ell_{BC_2C_3})}{\sinh(\omega \ell_{C_2})\sinh(\omega \ell)} \,.
\end{equation}

\medskip
\textit{Neumann Boundary conditions} --- Neumann Boundary conditions are achieved by adding a field $\psi$ at each boundary vertices and integrating over each individual $\psi$ as follows
\begin{equation}
\begin{aligned}
\mathcal{Z}_n^{(3)} &= \int d\phi_1 ... d\phi_{4n} \prod_{i=1}^{n} \bigg(\int d\psi K(\psi,\phi_{2n+2i-1};\ell_{C_1})\bigg)^n \bigg (\int d\psi K(\psi,\phi_{2n+2i};\ell_{C_3})\bigg)^n  \\
&  \hspace{2.5cm} \times\prod_{i=1}^{2n}K(\phi_{2n+2i-1},\phi_{2i-1};\ell_A)^nK(\phi_{2n+2i},\phi_{2i};\ell_B)^n \\
& \hspace{2.95cm}\times\prod_{j=1}^{2n-1} \hspace{-2pt}K(\phi_j,\phi_{j+1};\ell_{C_2})\hspace{-2pt}\prod_{k=1}^{\hspace{2pt}\big\lceil\spm \spm \frac{2n-j}{2} \spm\spm\big\rceil \spm-1}\hspace{-7pt}K(\phi_j,\phi_{j+2k+1};\ell_{C_2}) \\
&=  \Bigg(\frac{(\omega\epsilon/\pi)^3}{\sinh(\omega \ell_A) \sinh(\omega \ell_B) \cosh(\omega \ell_{C_1})\sinh(\omega \ell_{C_2})\cosh(\omega \ell_{C_3})} \Bigg)^{n^2/2} \sqrt{\frac{\pi^{4n}}{\det M_{n}^{(3)}}}\,,
\end{aligned}
\end{equation}
where $M_{n}^{(3)}$ has the same structure as for Dirichlet boundary conditions, with each $\coth(\omega\ell_{C_i})$ replaced by a $\tanh(\omega\ell_{C_i})$, $i=1,3$. The genuine multientropy is then
\begin{equation}
G_{N,n}^{(3)}(A:B:C)=\frac{2-n}{4n}\log \frac{\cosh(\omega \ell_{AC_1C_2})\cosh(\omega \ell_{BC_2C_3})}{\sinh(\omega \ell_{C_2})\sinh(\omega \ell)} \spp.
\end{equation}

\subsubsection{Adjacent $A, B$ in the bulk}

\begin{figure*}[ht]
\centering
\includegraphics[scale=1]{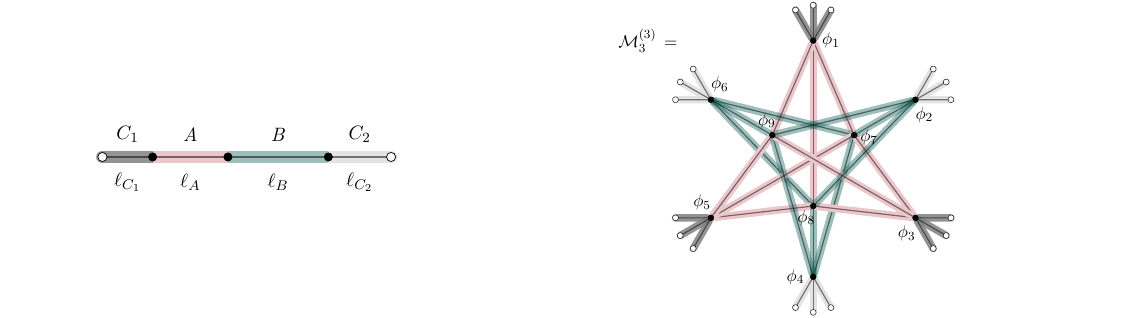}
\caption{Adjacent regions $A$ and $B$ in the bulk. Right: Replica graph resulting from the multientropy symmetry for $n=3$.}
\lb{AdjBulk}
\end{figure*}

\textit{Multientropy} --- From here on out, we only show the key quantities for the derivation of the (genuine) multientropies. For adjacent $A,B$ in the bulk of an interval, we draw the replica graph $\M_{3}^{(3)}$ in Fig.~\ref{AdjBulk}, which is straightforwardly generalized to arbitrary $n$ and yields
\begin{equation}
\begin{aligned}
\mathcal{Z}_n^{(3)} &=\int d\phi_1 ... d\phi_{3n} \prod_{i=1}^{n} \bigg(K(0,\phi_{2i-1};\ell_{C_1})^n K(0,\phi_{2i};\ell_{C_2})^n  \prod_{j=1}^n K(\phi_{2i-1},\phi_{2n+j};\ell_A)K(\phi_{2i},\phi_{2n+j};\ell_B) \bigg) \\
&= \Bigg(\frac{\omega^4}{\pi^4 \sinh(\omega \ell_A) \sinh(\omega \ell_B) \sinh(\omega \ell_{C_1})\sinh(\omega \ell_{C_2})} \Bigg)^{n^2/2} \sqrt{\frac{\pi^{3n}}{\det M_{n}^{(3)}}}\spp,
\end{aligned}
\end{equation}
where $M_{n}^{(3)}$ is  a $3n \times 3n$ matrix given in (\ref{MatAdj}), whose determinant evaluates to
\begin{equation}
    \det M_{n}^{(3)}=(n \omega)^{3n} \frac{\sinh( \omega \ell)\big(\spm \sinh(\omega \ell_{AB})\sinh(\omega \ell_{AC_1})\sinh(\omega \ell_{BC_2})\big)^{n-1}}{\big(\spm\sinh(\omega \ell_{C_1})\sinh(\omega \ell_{C_2})\big)^n \big(\spm\sinh(\omega \ell_A)\sinh(\omega \ell_B)\big)^{2n-1} } \, \,.
\end{equation}
The multientropy follows as
\begin{equation}
    \begin{aligned}
S_n^{(3)}(A:B:C)&=\log \frac{\sinh(\omega \ell_{C_1})^{\frac{1}{2}}\sinh(\omega \ell_{C_2})^{\frac{1}{2}}\sinh(\omega \ell_{AC_1})^{\frac{1}{2n}}\sinh(\omega \ell_{BC_2})^{\frac{1}{2n}}\sinh(\omega \ell_{AB})^{\frac{1}{2n}}}{\omega^\frac{3}{2} \sinh(\omega \ell_A)^{\frac{1-n}{2n}}\sinh(\omega \ell_B)^{\frac{1-n}{2n}}\sinh(\omega \ell)^{\frac{n+1}{2n}}}+ \frac{3}{2}\log \pi + \frac{3\log n}{2(n-1)} \spp.
\end{aligned}
\end{equation}

\medskip
\textit{Genuine multientropy} --- As previously, we give the averaged sum of Rényi entanglement entropies
\begin{equation}
    \begin{aligned}
\frac{1}{2}\big ( S_n(A)+S_n(B)+S_n(C) \big )&= \frac{1}{4} \log \frac{\sinh^2(\omega \ell_{C_1})\sinh(\omega \ell_A)\sinh(\omega \ell_B)\sinh^2(\omega \ell_{C_2})\sinh(\omega \ell_{AC_1})\sinh(\omega \ell_{AB})\sinh(\omega \ell_{BC_2})}{\omega^6 \sinh^3(\omega \ell)} \\
&\hspace{1.5cm} +\frac{3}{2}\log \pi + \frac{3\log n}{2(n-1)} \spp.
\end{aligned}
\end{equation}
Genuine multientropy is obtained by subtracting it from the multientropy
\begin{equation}
G_n^{(3)}(A:B:C)=\frac{2-n}{4n}\log \frac{\sinh(\omega \ell_{AB})\sinh(\omega \ell_{AC_1})\sinh(\omega \ell_{BC_2})}{\sinh(\omega \ell_{A})\sinh(\omega \ell_{B})\sinh(\omega \ell)} \spp.
\end{equation}

For Neumann boundary conditions, repeating the same process as in the previous case gives
\begin{equation}
G_{N,n}^{(3)}(A:B:C)=\frac{2-n}{4n}\log \frac{\sinh(\omega \ell_{AB})\cosh(\omega \ell_{AC_1})\cosh(\omega \ell_{BC_2})}{\sinh(\omega \ell_{A})\sinh(\omega \ell_{B})\sinh(\omega \ell)} \spp.
\end{equation}

\subsubsection{Disjoint $A, B$ with $B$ in the bulk}
\begin{figure*}[ht]
\centering
\includegraphics[scale=1]{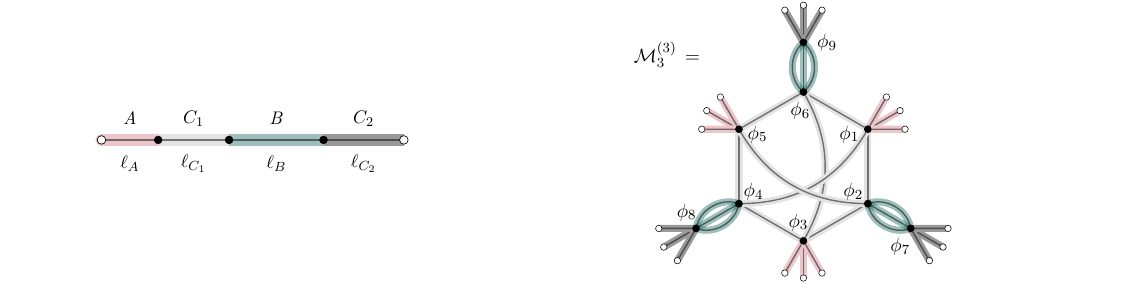}
\caption{Disjoint regions $A$ and $B$, with $A$ on the boundary and $B$ in the bulk. Right: Replica graph resulting from the multientropy symmetry for $n=3$.}
\lb{DisBoundBulk}
\end{figure*}

For the geometry shown in Fig.~\ref{DisBoundBulk}, together with the replica graph for $n=3$ which is easily generalized. We obtain the following partition function
\begin{equation}
    \begin{aligned}
\mathcal{Z}_n^{(3)} &= \int d\phi_1 ... d\phi_{3n} \prod_{i=1}^{n} K(0,\phi_{2i-1};\ell_A)^n \bigg (K(0,\phi_{2n+i};\ell_{C_2})K(\phi_{2n+i},\phi_{2i};\ell_B) \bigg)^n \\
& \hspace{2.5cm}\times\prod_{j=1}^{2n-1} \hspace{-2pt}K(\phi_j,\phi_{j+1};\ell_{C_1})\hspace{-2pt}\prod_{k=1}^{\hspace{2pt}\big\lceil\spm \spm \frac{2n-j}{2} \spm\spm\big\rceil \spm-1}\hspace{-7pt}K(\phi_j,\phi_{j+2k+1};\ell_{C_1}) \\
& = \Bigg(\frac{\omega^4}{\pi^4 \sinh(\omega \ell_A) \sinh(\omega \ell_B) \sinh(\omega \ell_{C_1})\sinh(\omega \ell_{C_2})} \Bigg)^{n^2/2} \sqrt{\frac{\pi^{3n}}{\det M_{n}^{(3)}}}\spp.
\end{aligned}
\end{equation}
$M_{n}^{(3)}$ is a $3n \times 3n$ matrix shown in (\ref{MatDis3}), with determinant given by
\begin{equation}
\det M_{n}^{(3)}=(n\omega)^{3n} \frac{\sinh(\omega \ell)\big(\spm\sinh(\omega \ell_{AC_1})\sinh(\omega \ell_{BC_1C_2})\big)^{n-1}}{\big(\spm\sinh(\omega \ell_{A})\sinh(\omega \ell_{B})\sinh(\omega \ell_{C_2}) \big)^n\sinh(\omega \ell_{C_1})^{2n-1}} \spp,
\end{equation}
from which follows the multientropy
\begin{equation}
S_n^{(3)}=\log \frac{\sinh(\omega \ell_{A})^\frac{1}{2}\sinh(\omega \ell_{B})^\frac{1}{2}\sinh(\omega \ell_{C_2})^\frac{1}{2}\sinh(\omega \ell_{AC_1})^\frac{1}{2n}\sinh(\omega \ell_{BC_1C_2})^\frac{1}{2n}}{\omega^{3/2}\sinh(\omega \ell_{C_1})^\frac{1-n}{2n}\sinh(\omega \ell)^\frac{n+1}{2n}}+\frac{3}{2}\log \pi +\frac{3\log n}{2(n-1)}\spp.
\end{equation}

\medskip
\textit{Genuine multientropy} --- The averaged sum of Rényi entanglement entropies reads
\begin{equation}
\begin{aligned}
\frac{1}{2}\big( S_n(A)+S_n(B)+S_n(C) \big)&=\frac{1}{4} \log \frac{\sinh^2(\omega \ell_{A})\sinh^2(\omega \ell_{B})\sinh^2(\omega \ell_{C_2})\sinh(\omega \ell_{C_1})\sinh(\omega \ell_{AC_1})\sinh(\omega \ell_{BC_1C_2})}{\omega^6 \sinh^3(\omega \ell)} \\
&\hspace{1.5cm}+\frac{3}{2}\log \pi + \frac{3\log n}{2(n-1)} \,, 
\end{aligned}
\end{equation}
which, subtracted from the multientropy, leads to the genuine multientropy
\begin{equation}
G_n^{(3)}(A:B:C)=\frac{2-n}{4n}\log \frac{\sinh(\omega \ell_{AC_1})\sinh(\omega \ell_{BC_1C_2})}{\sinh(\omega \ell_{C_1})\sinh(\omega \ell)}\spp .
\end{equation}

For Neumann boundary conditions we find
\begin{equation}
G_{N,n}^{(3)}(A:B:C)=\frac{2-n}{4n}\log \frac{\cosh(\omega \ell_{AC_1})\cosh(\omega \ell_{BC_1C_2})}{\sinh(\omega \ell_{C_1})\sinh(\omega \ell)}\spp.
\end{equation}

\subsubsection{Adjacent $A, B$ with $A$ on the boundary}

By symmetry of the genuine multientropy, from (\ref{GM_dis_DD1}) we obtain
\begin{equation}
G_{n}^{(3)}(A:B:C)=\frac{2-n}{4n}\log \frac{\sinh(\omega \ell_{AB}) \sinh(\omega \ell_{BC})}{\sinh(\omega\ell_B)\sinh(\omega \ell)}\spp,
\end{equation}
and similarly for Neumann boundary conditions
\begin{equation}
G_{N,n}^{(3)}(A:B:C)=\frac{2-n}{4n}\log \frac{\cosh(\omega \ell_{AB}) \cosh(\omega \ell_{BC})}{\sinh(\omega\ell_B)\sinh(\omega \ell)} \spp.
\end{equation}

\subsection{On a circle}

\subsubsection{Disjoint $A, B$}

\begin{figure*}[ht]
\centering
\includegraphics[]{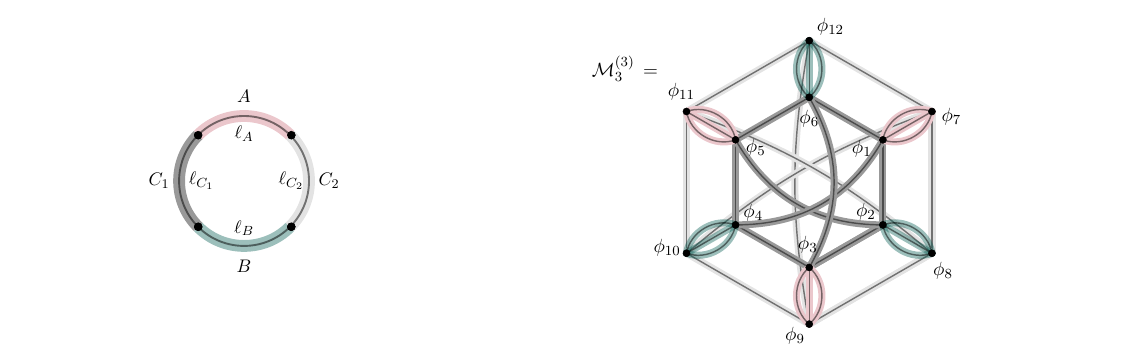}\vspace{-10pt}
\caption{Disjoint regions $A$ and $B$ on a circle. Right: Replica graph resulting from the multientropy symmetry for $n=3$.}
\lb{DisPeriodic}
\end{figure*}

\textit{Multientropy} --- For disconnected $A$ and $B$ on a circle (see Fig.\,\ref{DisPeriodic}), the partition function on the replica graph reads
\begin{equation}
\begin{aligned}
\mathcal{Z}_n^{(3)} &= \int d\phi_1 ... d\phi_{4n} \prod_{i=1}^{n} K(\phi_{2i-1},\phi_{2n+2i-1};\ell_A)^n K(\phi_{2i},\phi_{2n+2i};\ell_B)^n  \\
& \hspace{2.55cm}\times\prod_{j=1}^{2n-1} \hspace{-2pt}K(\phi_j,\phi_{j+1};\ell_{C_1})\hspace{-2pt}\prod_{k=1}^{\hspace{2pt}\big\lceil\spm \spm \frac{2n-j}{2} \spm\spm\big\rceil \spm-1}\hspace{-7pt}K(\phi_j,\phi_{j+2k+1};\ell_{C_1}) \\
&  \hspace{3.12cm} \times\prod_{l=1}^{2n-1} \hspace{-2pt}K(\phi_{2n+l},\phi_{2n+l+1};\ell_{C_2})\hspace{-2pt}\prod_{q=1}^{\hspace{2pt}\big\lceil\spm \spm \frac{2n-l}{2} \spm\spm\big\rceil \spm-1}\hspace{-7pt}K(\phi_{2n+l},\phi_{2n+l+2q+1};\ell_{C_2})\\
&= \Bigg(\frac{\omega^4}{\pi^4 \sinh(\omega \ell_A) \sinh(\omega \ell_B) \sinh(\omega \ell_{C_1})\sinh(\omega \ell_{C_2})} \Bigg)^{n^2/2} \sqrt{\frac{\pi^{4n}}{\det M_{n}^{(3)}}}\spp,
\end{aligned}
\end{equation}
where $M_{n}^{(3)}$ is a $4n \times 4n$ matrix given in (\ref{MatDis4}). Its determinant reads
\begin{equation}
\det M_{n}^{(3)}=4(n\omega)^{4n} \frac{\sinh(\omega \ell/2)^2 \big(\spm\sinh(\omega \ell_{AC_1C_2}) \sinh(\omega \ell_{BC_1C_2})\big)^{n-1}}{\big(\spm\sinh(\omega \ell_A)\sinh(\omega \ell_B)\big)^n\big(\spm\sinh(\omega \ell_{C_1})\sinh(\omega \ell_{C_2})\big)^{2n-1}}\spp.
\end{equation}
The multientropy is then found to be
\begin{equation}
\begin{aligned}
S^{(3)}_n(A:B:C)&=\log \frac{\sinh(\omega \ell_A)^{\frac{1}{2}}\sinh(\omega \ell_B)^{\frac{1}{2}}\sinh(\omega \ell_{AC_1C_2})^{\frac{1}{2n}}\sinh(\omega \ell_{BC_1C_2})^{\frac{1}{2n}}}{2^{\frac{n+1}{n}}\omega^2 \sinh(\omega \ell_{C_1})^{\frac{1-n}{2n}}\sinh(\omega \ell_{C_2})^{\frac{1-n}{2n}}\sinh(\omega \ell)^{\frac{n+1}{n}}}+ 2\log \pi + \frac{2\log n}{n-1} \spp.
\end{aligned}
\end{equation}

\medskip
\textit{Genuine multientropy} ---  The averaged sum of Rényi entanglement entropies is

\begin{equation}
\begin{aligned}
    \frac{1}{2}\big ( S_n(A)+S_n(B)+S_n(C) \big )=& \frac{1}{4}\log \frac{\sinh^2(\omega \ell_A)\sinh^2(\omega \ell_B)\sinh(\omega \ell_{C_1})\sinh(\omega \ell_{C_2})\sinh(\omega \ell_{AC_1C_2})\sinh(\omega \ell_{BC_1C_2})}{64 \omega^8 \sinh^6(\omega \ell/2)}\\
    & \hspace{1cm} + 2\log \pi + \frac{2\log n}{n-1} \spp ,
\end{aligned}
\end{equation}
subtracting this quantity from the multientropy leads to
\begin{equation}
    G_n^{(3)}(A:B:C)=\frac{2-n}{4n}\log \frac{\sinh(\omega \ell_{AC_1C_2})\sinh(\omega \ell_{BC_1C_2})}{4\sinh(\omega \ell_{C_1})\sinh(\omega \ell_{C_2})\sinh^2(\omega \ell/2)} \spp.
\end{equation}

\subsubsection{Adjacent $A, B$}

\begin{figure*}[ht]
\centering
\includegraphics[]{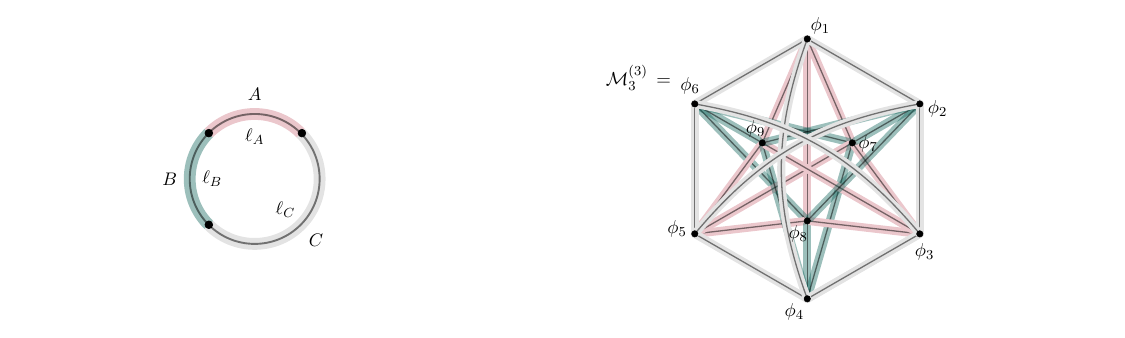}\vspace{-10pt}
\caption{Adjacent regions $A$ and $B$ on a circle. Right: Replica graph resulting from the multientropy symmetry for $n=3$.}
\lb{AdjPeriodic}
\end{figure*}

\textit{Multientropy} --- The replica graph yields the partition function
\begin{equation}
\begin{aligned}
\mathcal{Z}_n^{(3)} &= \int d\phi_1 ... d\phi_{3n} \prod_{i,j=1}^{n}K(\phi_{2i-1},\phi_{2n+j};\ell_A)K(\phi_{2i},\phi_{2
n+j};\ell_B) \\
& \hspace{2.65cm} \times\prod_{j=1}^{2n-1} \hspace{-2pt}K(\phi_j,\phi_{j+1};\ell_{C})\hspace{-2pt}\prod_{k=1}^{\hspace{2pt}\big\lceil\spm \spm \frac{2n-j}{2} \spm\spm\big\rceil \spm-1}\hspace{-7pt}K(\phi_j,\phi_{j+2k+1};\ell_{C}) \\
&= \Bigg(\frac{\omega^3}{\pi^3 \sinh(\omega \ell_A) \sinh(\omega \ell_B) \sinh(\omega \ell_{C})} \Bigg)^{n^2/2} \sqrt{\frac{\pi^{3n}}{\det M_{n}^{(3)}}} \spp ,
\end{aligned}
\end{equation}
$M_{n}^{(3)}$ is a $3n \times 3n$ matrix shown in (\ref{MatAdj2}), for which we derive the following determinant for all $n$
\begin{equation}
\det M_{n}^{(3)} = 4 (n\omega)^{3n}  \frac{\sinh^2(\omega \ell/2)\big(\spm \sinh(\omega \ell_{AB})\sinh(\omega \ell_{BC})\sinh(\omega \ell_{AC})\big)^{n-1}}{\big(\spm\sinh(\omega \ell_A)\sinh(\omega \ell_B)\sinh(\omega \ell_C)\big)^{2n-1}}\spp.
\end{equation}
The multientropy follows
\begin{equation}
    \begin{aligned}
S_n^{(3)}(A:B:C)= \log \frac{\sinh(\omega \ell_{AB})^\frac{1}{2n}\sinh(\omega \ell_{BC})^\frac{1}{2n}\sinh(\omega \ell_{AC})^\frac{1}{2n}}{2^\frac{n+1}{n}\omega^\frac{3}{2}\sinh(\omega \ell_A)^\frac{1-n}{2n}\sinh(\omega \ell_B)^\frac{1-n}{2n}\sinh(\omega \ell_C)^\frac{1-n}{2n}\sinh(\omega \ell/2)^\frac{n+1}{n}}+\frac{3}{2}\log \pi + \frac{3\log n}{2(n-1)} \spp.
\end{aligned}
\end{equation}

\medskip
\textit{Genuine multientropy} --- The averaged sum of Rényi entanglement entropies reads
\begin{equation}
\begin{aligned}
\frac{1}{2}\big ( S_n(A)+S_n(B)+S_n(C) \big ) &= \frac{1}{4} \log \frac{\sinh(\omega \ell_A)\sinh(\omega \ell_B)\sinh(\omega \ell_C)\sinh(\omega \ell_{AB})\sinh(\omega \ell_{BC})\sinh(\omega \ell_{AC})}{64 \omega^6 \sinh^6(\omega \ell/2)}\\
& \hspace{1.5cm}+\frac{3}{2}\log \pi+\frac{3\log n}{2(n-1)} \spp,
\end{aligned}
\end{equation}
and the genuine multientropy is obtained from subtracting this sum from the multientropy
\begin{equation}
G_n^{(3)}(A:B:C)=\frac{2-n}{4n}\log \frac{\sinh(\omega \ell_{AB})\sinh(\omega \ell_{BC})\sinh(\omega \ell_{AC})}{4\sinh(\omega \ell_A)\sinh(\omega \ell_B)\sinh(\omega \ell_C) \sinh^2(\omega \ell/2)}\spp .
\end{equation}

\section{Dihedral invariants}\label{dihedral}

In this appendix, we discuss an alternative definition of dihedral invariant involving a different normalization. We also compute the dihedral invariant \eqref{di} for Lifshitz groundstates, GHZ and W states, and check it is equal to the $m=2$ reflected entropy, in accordance with \eqref{rel_di_SR}.

\subsection{Alternative normalization and definition of dihedral invariant}\lb{Apdx_norm_di}

In defining the family of dihedral invariants \eqref{di}, we chose a particular prescription for normalization. Though our choice appears natural since we prove that the dihedral invariant \eqref{di} is exactly the $(2,n)$--Rényi reflected entropy, see \eqref{rel_di_SR}, it is interesting to explore other normalizations.

We introduce the family of dihedral invariants with normalization labeled by a parameter $q>0$,
\bea
\D_{2n}^{(q)}(A:B)=\frac{1}{1-qn}\log \frac{\mathcal{Z}_{2n}}{(\mathcal{Z}_{2/q})^{qn}} \,,
\lb{norm}
\eea
such that the limit $n\rightarrow 1/q$ is regular. 
Using the relations $\log\Z_{2n} = \E_n^{\scalebox{0.5}{\rm CCNR}}(A:C)=(1-n)S_{2,n}^R(A:C)-nS_2(A\cup C)$ (see \eqref{di-ccnr} for the first equality and \cite{Berthiere:2023gkx} for the second), the dihedral invariants defined in \eqref{norm} can be written as
\bea
\begin{split}
 \D_{2n}^{(q)}(A:B) &=\frac{\E_n^{\scalebox{0.5}{\rm CCNR}}(A:C)-qn\spp \E_{1/q}^{\scalebox{0.5}{\rm CCNR}}(A:C)}{1-qn} \\
    &= \frac{(1-n)S_{2,n}^R(A:C)-n(q-1)S^R_{2,1/q}(A:C)}{1-qn} \,.
    \lb{ZtoSR}
\end{split}
\eea
Taking the limit $n\rightarrow 1/q$ in \eqref{ZtoSR} yields
\bea
\begin{split}
    \D^{(q)}(A:B)\equiv\lim_{n\rightarrow 1/q}\D_{2n}^{(q)}(A:B) &=\E_{1/q}^{\scalebox{0.5}{\rm CCNR}}(A:C)-\frac{1}{q} \partial_n \E_n^{\scalebox{0.5}{\rm CCNR}}(A:C)\Big|_{n=1/q} \\
    &= S^R_{2,1/q}(A:C)-\frac{q-1}{q^2} \partial_n S^R_{2,n}(A:C) \Big|_{n=1/q} \,\spp.
\end{split}
\eea
We recover relation \eqref{rel_di_SR} for $q=1$. As an example, choosing $q=2$ leads to
\begin{equation}
    \mathcal{D}^{(q=2)}(A:B)=S^R_{2,1/2}(A:C) -\frac{1}{4}\partial_nS^R_{2,n}(A:C) \big|_{n=1/2} \,\,.
\end{equation}

\subsection{Dihedral invariant for pure tripartite qubit states}

We compute the dihedral invariant and reflected entropy of generalized GHZ and W states, and show that these two quantities are equivalent.

\subsubsection{GHZ states}

Consider the generalized GHZ state defined as
\begin{equation}
    |{\rm GHZ}(\theta)\rangle = \cos \theta\spp |000\rangle + \sin \theta\spp |111 \rangle \,.
\end{equation}
Since this state is symmetric under qubit permutations, the dihedral invariant is straightforward to compute~\cite{Gadde:2022cqi}
\begin{equation}
    \mathcal{D}_{2n}(\theta)= \frac{1}{1-n} \log \frac{\cos^{4n} \theta+\sin^{4n}\theta}{\big(\spm\cos^4\theta + \sin^4 \theta \big)^n}\,.
\end{equation}
%
%
%
The reflected entropy is also easily calculated. We start with $\rho_{AC}$
\begin{equation}
    \rho_{AC}=\cos^2 \theta |00\rangle\langle00|+\sin^2 \theta |11\rangle\langle 11|\,,
\end{equation}
the reflected density matrix $\rho_{AA}^{(m)}$ then follows as
\begin{equation}
    \rho_{AA}^{(m)}=\frac{1}{\cos^4 \theta + \sin^4 \theta}\big( \spm\cos^4 \theta |00\rangle\langle00|+\sin^4 \theta |11\rangle\langle 11 |\big) \,,
\end{equation}
and the reflected entropy reads
\begin{equation}
    S^R_{2,n}(\theta) =\frac{1}{1-n}\log \tr \big(\rho_{AA}^{(m)}\big)^n= \frac{1}{1-n} \log \frac{\cos^{4n} \theta+\sin^{4n}\theta}{\big(\spm\cos^4\theta + \sin^4 \theta \big)^n}\,.
\end{equation}
Both quantities thus match, as expected from \eqref{rel_di_SR}.

\pagebreak
\subsubsection{W state}

The generalized W state is defined as
\begin{equation}
 |W(\theta,\varphi)\rangle = \cos \theta |001\rangle + \sin \theta \cos \varphi |010\rangle + \sin \theta \sin \varphi |100\rangle \,. 
\end{equation}
The dihedral invariants and reflected entropies are derived for $n=2,3$ from a direct evaluation of the contractions of the density matrices with their respective permutations. It can be checked that
\begin{equation}
    \mathcal{D}_4(A:B)=2\log \frac{4\cos^4\theta + 8\cos^2 \theta \sin^2 \theta \sin^2 \varphi+(3+\cos 4\varphi)\sin^4 \theta}{ 4 \cos^4 \theta + (3+\cos 4\varphi )\sin^4\theta}= S^R_{2,2}(A:C) \,\spp,
\end{equation}
and
\begin{equation}
    \mathcal{D}_6(A:B)=\log \frac{4\cos^4\theta + 8\cos^2 \theta \sin^2 \theta \sin^2 \varphi+(3+\cos 4\varphi)\sin^4 \theta}{4\cos^4\theta - 4\cos^2 \theta \sin^2 \theta \sin^2 \varphi+(3+\cos 4\varphi)\sin^4 \theta}= S^R_{2,3}(A:C)\, \spp .
\end{equation}
Note that we recover the results found in \cite{Harper:2025uui} for both GHZ and W states with their choice of normalization.

\subsection{Dihedral invariant for Lifshitz groundstates}
\subsubsection{Disjoint regions on the boundary}

\begin{figure*}[ht]
\centering
\includegraphics[scale=0.8]{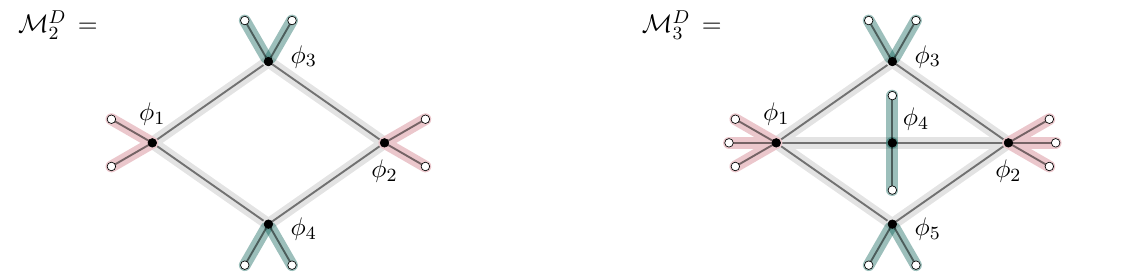}
\caption{Replica graphs for the dihedral invariants of two disjoint regions $A, B$ on the boundary (see Fig.\,\hyperref[fig_tri]{\ref{fig_tri}(a)}), for $n=2,3$.}
\lb{DihedralDis}
\end{figure*}

For two disjoint regions $A,B$, shown in Fig.\,\hyperref[fig_tri]{\ref{fig_tri}(a)}, we illustrates the replica graph for two different $n$ in Fig.\,\ref{DihedralDis}, which generalizes naturally. The partition function reads
\begin{equation}
    \begin{aligned}
        \mathcal{Z}_{2n}&= \int d\phi_1...d\phi_{n+2} \prod_{i=1}^2K(0,\phi_i;\ell_A)^n\prod_{j=3}^{n+2}  K(\phi_i,\phi_j;\ell_C)K(0,\phi_j;\ell_B)^2 \\
        &= \bigg(\frac{\omega^3\epsilon^3}{\pi^3\sinh(\omega \ell_A)\sinh(\omega \ell_B)\sinh(\omega \ell_C)}\bigg)^n \sqrt{\frac{\pi^{n+2}}{\det M_n^{(D)}}}\spp,
    \end{aligned}
\end{equation}
where $M_n^{(D)}$ is a $(n+2)\times (n+2)$ matrix presented in \eqref{DihedralMatDis}. Its determinant reads
\begin{equation}
    \det M_n^{(D)}=2^n (\omega\epsilon)^{n+2} n^2 \frac{\sinh(\omega \ell_{BC})^{n-1}\sinh(\omega \ell_{AC})\sinh(\omega \ell)}{\sinh(\omega \ell_B)^n\sinh(\omega \ell_C)^{n+1}\sinh^2(\omega \ell_A)} \,.
\end{equation}
The dihedral invariant is then evaluated using \eqref{di}
\begin{equation}
    \mathcal{D}_{2n}(A:B)=\frac{1}{2}\log \frac{\sinh^2(\omega \ell_A) \sinh(\omega \ell_C)\sinh(\omega \ell_{BC})}{\omega^2 \epsilon^2 \sinh(\omega \ell_{AC})\sinh(\omega \ell)}+\frac{\log n}{n-1} +\log \pi  = S_{2,n}^R(A:C)= I_n(A:C) \spp.
\end{equation}
It was shown in \cite{Berthiere:2023bwn} that the $m=2$ reflected entropy is equal to the mutual information.

\subsubsection{Adjacent regions on the boundary}
\vspace{-0.2cm}
\begin{figure*}[ht]
\centering
\hspace*{-1cm}
    \begin{minipage}[c]{0.48\textwidth}
        \centering
        \includegraphics[scale=1.5,trim=100 0 100 0, clip]{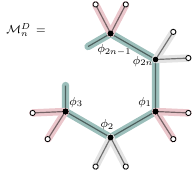}
    \end{minipage}
\caption{Replica graph for the dihedral invariants of two adjacent regions $A$ and $B$, $B$ being in the bulk and $A$ on the boundary.}
\lb{DihedralAdj}
\end{figure*}

The replica graph Fig.\,\ref{DihedralAdj} yields the partition function
\begin{equation}
    \begin{aligned}
        \mathcal{Z}_{2n}&= \int d\phi_1...d\phi_{2n} \prod_{i=1}^n K(0,\phi_{2i-1};\ell_A)^2  K(0,\phi_{2i};\ell_C)^2K(\phi_1,\phi_{2n};\ell_B) \prod_{j=1}^{2n-1}K(\phi_j,\phi_{j+1};\ell_B)\\
        &= \bigg(\frac{\omega^3\epsilon^3}{\pi^3\sinh(\omega \ell_A)\sinh(\omega \ell_B)\sinh(\omega \ell_C)}\bigg)^n \sqrt{\frac{\pi^{2n}}{\det M_n^{(D)}}} \spp
    \end{aligned}
\end{equation}
The $2n \times 2n$ matrix $M_n^{(D)}$ is shown in \eqref{DihedralMatAdj} and has determinant
\begin{equation}
    \det M_n^{(D)}=4 c^{2n}\sinh^2 n\theta \,, \qquad \theta=\text{arccosh}\sqrt{\eta^{-1}} \,,
\end{equation}
where
\begin{equation}
    \eta=\frac{\sinh(\omega \ell_A)\sinh(\omega \ell_C)}{\sinh(\omega \ell_{AB})\sinh(\omega \ell_{BC})} \,, \qquad c=\frac{-\omega}{\sinh(\omega \ell_B)} \spp.
\end{equation}
For this configuration, the dihedral invariant reads

\begin{equation}
\begin{aligned}
    \mathcal{D}_{2n}(A:B)=\frac{1}{n-1}\log \frac{(\sqrt{1-\eta}+1)^{2n}-\eta^n}{\big((\sqrt{1-\eta}+1)^2-\eta \big)^n}
    = S^R_{m,n}(A:C)\spp.
\end{aligned}
\end{equation}
In this situation, $S_{m,n}^R$ is identical for all $m$, see \cite{Berthiere:2023bwn}.

\section{Matrices and determinants}
\lb{Apdx:det}

For Lifshitz groundstates, partition functions $\Z(\M)$ on replica graphs $\M$ are determinants of certain matrices $M$,
\bea
\Z(\M)\propto\( \det M\)^{-1/2}.
\eea
We list here such matrices appearing in our calculations for the different cases we consider, as well as their determinant.
\medskip

\subsection{Multientropy}

\subsubsection{Disjoint regions on the boundary}

Associated to the replica graph shown in Fig.~\ref{fig_GDD} is the following $2n \times 2n$ matrix
\begin{equation}
M_n^{(3)}=\begin{pmatrix}
a & c & 0 & c & \cdots &    &  \\
c & b & c & 0 & c & \cdots  &  \\
& c & a & c & 0 &  c & \cdots \\[-0.1cm]
\ddots &  & \ddots & \ddots & \ddots & & \ddots  \\
\cdots &c &0 &c & b &c&0\\ 
 & \cdots& c  & 0 & c & a & c \\
 &  & \cdots & c& 0 & c & b \\
\end{pmatrix} ,
\label{MatDis}
\end{equation}
where $a=n \omega \big(\spm\coth(\omega \ell_A) + \coth(\omega \ell_C) \big)$, $b=n \omega \big(\spm\coth(\omega \ell_B) + \coth(\omega \ell_C) \big)$ and $c=-\omega/\sinh(\omega \ell_C)$.

\medskip
\textit{Determinant} --- We begin by modifying the matrix $M_n^{(3)}$. Since there is an equal number of $a$'s and $b$'s on the diagonal, replacing each instance with $\sqrt{ab}$ doesn't modify the determinant. This results in a circulant matrix with $c_0=\sqrt{ab}$, $c_{2i}=0$ and $c_{2i-1}=c,0<i\leq n$. The identity for computing the determinant of a circulant matrix yields
\begin{equation}
\begin{aligned}
\det M_n^{(3)}&=\prod_{j=0}^{2n-1}\bigg(c_0+c_{2n-1}\omega^j+c_{2n-2}\omega^{2j} + \hdots + c_1\omega^{(2n-1)j} \bigg)  , \quad\omega = \exp\Big(\frac{2\pi i}{2n} \Big)\\
&=  \prod_{j=0}^{2n-1} \bigg(\sqrt{ab} + c\sum_{k=1}^{n} \omega^{(2k-1)j} \bigg)\\
&=  \prod_{j=0}^{2n-1} \bigg(\sqrt{ab} + c\omega^{-j}\sum_{k=1}^{n} \omega^{2kj} \bigg) \spp.
\end{aligned}
\end{equation}
The geometric sum can be evaluated when $\omega^{2kj} \neq 1$, which is the case for $j\neq 0,n$, as follows
\begin{equation}
\sum_{k=1}^{n} (\omega^{2j})^k=\frac{1-\omega^{2nj}}{1-\omega^{2j}}=0\,, \qquad  (\omega^{2nj}=1, \,\forall j\in \mathbb{Z}) \,.
\end{equation}
For $j=0,n$; $\omega^{2kj}=1$ and $\omega^{-j}=\pm1$, hence
\begin{equation}
\det M_n^{(3)}=\big(\sqrt{ab} + nc \big)\big(\sqrt{ab} - nc \big) \prod_{j=1,j\neq n}^{2n-1} \sqrt{ab} = (ab)^{n-1} (ab-n^2c^2) \spp.
    \label{Mat1Det}
\end{equation}
Changing $a,b$ and $c$ to their respective values and using the identities
\begin{equation}
\coth(x)+\coth(y) = \frac{\sinh(x+y)}{\sinh(x)\sinh(y)}\,,
\end{equation}
and
\begin{equation}
\frac{\sinh(x+y)\sinh(y+z)}{\sinh(x)\sinh^2(y)\sinh(z)}-\frac{1}{\sinh^2(y)}=\frac{\sinh(x+y+z)}{\sinh(x)\sinh(y)\sinh(z)}\spp .
\end{equation}
finalizes the derivation.

\subsubsection{Disjoint regions in the bulk}

For this configuration, see Fig.~\ref{DisBulk}, the corresponding $M_n^{(3)}$ is a $4n\times 4n$ block matrix of the form
\begin{equation}
M_n^{(3)}=\begin{pmatrix}
X & Y \\
Y & Z
\end{pmatrix}, \qquad X,Y,Z\in \mathcal{M}_{2n}(\mathbb{R}) \,,
\label{MatDis2}
\end{equation}
where $X$ has the same structure as \eqref{MatDis} with $a=n \omega \big(\spm\coth(\omega \ell_A) + \coth(\omega \ell_{C_2}) \big)$, $b=n \omega \big(\spm\coth(\omega \ell_B) + \coth(\omega \ell_{C_2}) \big)$ and $c=-\omega/\sinh(\omega \ell_{C_2})$. $Y$ and $Z$ read
\begin{equation}
Y=\text{diag} \big(\alpha,\beta,\alpha,\beta,...  \big)\,,\qquad 
    \alpha=\frac{-n\omega}{\sinh(\omega \ell_A)}\,, \quad \beta= \frac{-n\omega}{\sinh(\omega \ell_B)}\,.
\end{equation}
\begin{equation}
    Z=\text{diag} \big (\gamma_1,\gamma_2,\gamma_1,\gamma_2,...  \big)\,,\qquad  \begin{matrix}
    \gamma_1=n \omega \big(\spm\coth(\omega \ell_A) + \coth(\omega \ell_{C_1}) \big)\,, \\[1ex]
    \gamma_2=n \omega \big(\spm\coth(\omega \ell_B) + \coth(\omega \ell_{C_3}) \big)\,.
\end{matrix}
\end{equation}

\medskip
\textit{Determinant} --- We make use of the following identity for block matrices
\begin{equation}
\det M_n^{(3)}=\det(Z)\det(X-YZ^{-1}Y)=\det(Z)\det(\tilde{X})\spp, 
\end{equation}
where $\tilde{X}$ has the same structure as $X$ with $\tilde{a}=a-\frac{\alpha^2}{\gamma_1}$, $\tilde{b}=b-\frac{\beta^2}{\gamma_2}$, and $\tilde{c}=c$. Using \eqref{Mat1Det}, we obtain
\bea
\det M_n^{(3)}&=(\gamma_1\gamma_2)^n\bigg(a-\frac{\alpha^2}{\gamma_1}\bigg)^{n-1}\bigg(b-\frac{\beta^2}{\gamma_2}\bigg)^{n-1} \bigg[\bigg(a-\frac{\alpha^2}{\gamma_1}\bigg)\bigg(b-\frac{\beta^2}{\gamma_2}\bigg)-n^2c^2 \bigg]\nn \\[1ex]
&= \big(a\gamma_1-\alpha^2\big)^{n-1}\big(b\gamma_2-\beta^2\big)^{n-1} \bigg(\big(a\gamma_1-\alpha^2\big)\big(b\gamma_2-\beta^2\big)-n^2c^2\gamma_1\gamma_2 \bigg)\spp.
    \label{Mat2Det}
\eea
Changing the variables to their respective value and performing simplifications yields the final result.

\subsubsection{Adjacent regions in the bulk}

The replica graph corresponding to two adjacent regions in the bulk is shown Fig.~\ref{AdjBulk} and the corresponding $M_n^{(3)}$ is a $3n\times 3n$ block matrix of the form
\begin{equation}
M_n^{(3)}=\begin{pmatrix}
\hspace{-0.2cm}X & \hspace{-0.2cm}Y \\
Y^T & \hspace{-0.2cm}Z
\end{pmatrix}, \quad X \in \mathcal{M}_{2n}(\mathbb{R})\,, \;\; Y \in \mathcal{M}_{2n,n}(\mathbb{R})\,, \;\; Z \in \mathcal{M}_{n}(\mathbb{R})\, ,
\label{MatAdj}
\end{equation}
where
\begin{equation}
X=\text{diag} \big(a,b,a,b,... \big) \,, \quad \begin{matrix}
    a=n \omega \big (\spm \coth(\omega \ell_A) + \coth(\omega \ell_{C_1}) \big)\,, \\[1ex]
    b=n \omega \big ( \spm\coth(\omega \ell_B) + \coth(\omega \ell_{C_2}) \big)\,,
    \end{matrix}
\end{equation}
\begin{equation}
Y=\begin{pmatrix}
\alpha & \alpha & \cdots & \alpha\\
\beta & \beta & \cdots & \beta \\
\alpha & \alpha & \cdots & \alpha \vspace{-0.1cm}\\
\vdots & \vdots & \cdots & \vdots\\
\beta & \beta & \cdots & \beta\\
\end{pmatrix}, \qquad \alpha = \frac{-\omega}{\sinh(\omega \ell_A)} \, , \quad \beta= \frac{-\omega}{\sinh(\omega \ell_B)},
\label{Y}
\end{equation}
\begin{equation}
Z=\text{diag}(\gamma) \,,\qquad \qquad \hspace{0.6cm}\gamma=n\omega \big(\spm\coth( \omega \ell_A) + \coth (\omega \ell_B) \big) \,.
\label{Z}
\end{equation}

\medskip
\textit{Determinant ---} We begin by reindexing every $\phi$ such that $X=\text{diag}(a,a,...,a,b,b,...,b)$ and a $Y$ matrix with first all of the rows of $\alpha$s and then all of the rows of $\beta$s. We then compute the determinant using the identity
\begin{equation}
\det M_n^{(3)}=\det(Z)\det(X-YZ^{-1}Y^T)\,.
\end{equation} 
We start by computing $\det(X-YZ^{-1}Y^T)$
\begin{equation}
\begin{aligned}
\det(X-YZ^{-1}Y^T)&= \det\spm\spm\bigg[\begin{pmatrix}
aI_{n} & 0 \\ 0 & bI_{n}
\end{pmatrix}-\frac{1}{\gamma}\begin{pmatrix}
n \alpha^2 J_n & n\alpha \beta J_n \\
n \alpha \beta J_n &  n \beta^2J_n
\end{pmatrix} \bigg]\\ 
&= \frac{1}{\gamma^{2n}} \det \begin{pmatrix}
a\gamma I_n - n \alpha^2 J_n & - n\alpha \beta J_n \\
- n \alpha \beta J_n & b\gamma I_n-n \beta^2J_n
\end{pmatrix} \equiv \det \spm\spm \begin{pmatrix}
P_1 & P_2 \\ P_2 & P_3
\end{pmatrix}\spm,
\end{aligned}
\end{equation}
where $I_n$ is the $n\times n$ identity matrix and $J_n$ is the $n\times n$ matrix filled with ones. This is computed as
\begin{equation}
\det \spm\spm \begin{pmatrix}
P_1 & P_2 \\ P_2 & P_3
\end{pmatrix} = \det (P_1P_3-P_2^2)\spp.
\end{equation}
Using $J_n.J_n =n J_n$ and $\det(aJ_n+bI_n)=(an+b)b^{n-1}\,$, we find
\begin{equation}
\det M_n^{(3)} = (ab\gamma)^{n-1} \big (ab\gamma -n^2(a\beta^2+b\alpha^2)\big)\spp .
    \label{Mat3Det}
\end{equation}

\subsubsection{Disjoint regions with one at the boundary}

For this situation, the replica graph is presented Fig.~\ref{DisBoundBulk} and the associated $M_n^{(3)}$ reads
\begin{equation}
M_n^{(3)}=\begin{pmatrix}
X & Y \\
Y^T & Z
\end{pmatrix}, \quad X \in \mathcal{M}_{2n}(\mathbb{R})\,, \;\; Y \in \mathcal{M}_{2n,n}(\mathbb{R})\,, \;\; Z \in \mathcal{M}_{n}(\mathbb{R})\, ,
\label{MatDis3}
\end{equation}
$X$ has the same structure of \eqref{MatDis} with $a=n \omega \big(\spm\coth(\omega \ell_A) + \coth(\omega \ell_{C_1}) \big)$, $b=n \omega \big(\spm\coth(\omega \ell_B) + \coth(\omega \ell_{C_1}) \big)$ and \mbox{$c=-\omega/\sinh(\omega \ell_{C_1})$,} and
\begin{equation}
Y=\begin{pmatrix}
0 & 0 & 0 &\cdots&0\\
\beta & 0 & 0 & \cdots&0 \\
0 & 0 & 0 & \cdots&0 \\
0 & \beta & 0 & \cdots&0 \\
0 & 0 & 0 & \cdots & 0 \\
0 & 0 & \beta & \cdots & 0 \\
\vdots & \vdots & \vdots & & \vdots \\
0 & 0 & 0 & \cdots & \beta \\
\end{pmatrix}, \qquad \beta = \frac{-n\omega}{\sinh(\omega \ell_B)} \,,
\end{equation}
\begin{equation}
Z=\text{diag} (\gamma) \,, \qquad  \gamma=n\omega \big (\spm \coth(\omega \ell_B) + \coth (\omega \ell_{C_2}) \big) \,.
\end{equation}

\medskip
\textit{Determinant} --- %
$YZ^{-1}Y^T$ is a $2n \times 2n$ diagonal matrix with even diagonal entries filled with $\beta^2/\gamma$ and the rest being $0$. Thus, $\tilde{X}\equiv X-YZ^{-1}Y^T$  is of the form \eqref{MatDis} with $\tilde{a}=a, \tilde{b}=(b-\beta^2/\gamma)$ and $\tilde{c}=c$. Using \eqref{Mat1Det} we find
\begin{equation}
\begin{aligned}
\det M_n^{(3)}&=\det(Z)\det(\tilde{X}) \\
&=\gamma^n a^{n-1}\bigg(\spm b - \frac{\beta^2}{\gamma}\bigg)^{n-1}\bigg[a \bigg(b - \frac{\beta^2}{\gamma}\bigg) -n^2c^2 \bigg] \\
&= a^{n-1}\big(b\gamma - \beta^2\big)^{n-1}\big(a (b\gamma - \beta^2) -n^2c^2\gamma \big).
\end{aligned}
\label{Mat4Det}
\end{equation}

\subsubsection{Disjoint regions on a circle}

For two disjoint regions on a circle, the replica graph is shown Fig.~\ref{DisPeriodic} and the associated $M_n^{(3)}$ is a $4n \times 4n$ block matrix of the form
\begin{equation}
M_n^{(3)}=\begin{pmatrix}
X_1 & Y \\
Y & X_2
\end{pmatrix}, \qquad X_1,X_2,Y\in \mathcal{M}_{2n}(\mathbb{R}) \, ,
\label{MatDis4}
\end{equation}
where the $X_i$s have the same structure as \eqref{MatDis} with $a_i=n \omega \big(\spm\coth(\omega \ell_A) + \coth(\omega \ell_{C_i}) \big)$, $b_i=n \omega \big(\spm\coth(\omega \ell_B) + \coth(\omega \ell_{C_i}) \big)$ and $c_i=-\omega/\sinh(\omega \ell_{C_i})$, and
\begin{equation}
Y=\text{diag} (\alpha,\beta,\alpha,\beta,...) \,, \qquad  \alpha= \frac{-n \omega}{\sinh(\omega \ell_A)}\,, \quad \beta=\frac{-n \omega}{\sinh(\omega \ell_B)} \, .
\end{equation}

\medskip
\textit{Determinant ---} We reindex the $\phi$s such that
\begin{equation}
    M_n^{(3)}=\begin{pmatrix}
        X & Y \\ Y & Z
    \end{pmatrix}
\end{equation}
with 
\begin{equation}
    X=\begin{pmatrix}
      a_1I_n & \alpha I_n \\ \alpha I_n & a_2 I_n
    \end{pmatrix},\;\quad Y=\begin{pmatrix}
    c_1 J_n & 0 \\ 0 & c_2 J_n \end{pmatrix}, \quad\;  Z=\begin{pmatrix}
      b_1I_n & \beta I_n \\ \beta I_n & b_2 I_n
\end{pmatrix} .
\end{equation}
The determinant is computed with the following steps
\begin{equation}
    \det M_n^{(3)}=\det(X)\det(Z-YX^{-1}Y)=\det(X)\det \begin{pmatrix}
        P_1 & P_2 \\ P_2 & P_3 
    \end{pmatrix} = \det(X)\det(P_1)\det(P_3-P_2P_1^{-1}P_2) \spp.
\end{equation}
Similarly to adjacent regions in the bulk. The final result reads
\begin{equation}
\begin{aligned}
   \det M_n^{(3)}&=\Big(\spm\big(n^2c_1c_2-\alpha \beta\big)^2-n^2\big(a_2b_2c_1^2+a_1b_1c_2^2\big)-a_1a_2\beta^2-b_1b_2\alpha^2+a_1a_2b_1b_2\Big)\\
   & \hspace{.5cm}\times(a_1 a_2-\alpha^2)^{n-1}(b_1b_2-\beta^2)^{n-1} \spp.
\end{aligned}
\label{Mat5Det}
\end{equation}

\subsubsection{Adjacent regions on a circle}

This configuration and corresponding replica graph are shown Fig.~\ref{AdjPeriodic}, the related $M_n^{(3)}$ is the $3n\times 3n$ block matrix 
\begin{equation}
M_n^{(3)}=\begin{pmatrix}
\hspace{-0.2cm} X & \hspace{-0.2cm}Y \\
Y^T &\hspace{-0.22cm} Z
\end{pmatrix}, \quad X \in \mathcal{M}_{2n}(\mathbb{R})\,, \;\; Y \in \mathcal{M}_{2n,n}(\mathbb{R})\,, \;\; Z \in \mathcal{M}_{n}(\mathbb{R})\, ,
\label{MatAdj2}
\end{equation}
where $X$ is \eqref{MatDis}, $Y$ is \eqref{Y} and $Z$ is \eqref{Z}.

\medskip
\textit{Determinant ---} Its determinant is computed similarly to that for adjacent regions in the bulk, finding
\begin{equation}
    \det M_n^{(3)}=(ab\gamma)^{n-1}\big(2n^3 c\alpha \beta +ab\gamma-n^2(a\beta^2+b\alpha^2+c^2\gamma)  \big) \,\spp.
    \label{Mat6Det}
\end{equation}

\subsection{Dihedral invariant}

\subsubsection{Disjoint regions on the boundary}

The matrix $M_n^{(D)}$ associated with the partition function of the dihedral invariant for this situation, see Fig.\,\ref{DihedralDis}, is
\begin{equation}
    M_n^{(D)}=\begin{pmatrix}
        a & 0 & -c & \cdots  & \cdots & -c \\
        0 & a & -c &  \cdots & \cdots & -c \\
        -c & -c & b & 0 & \cdots&0 \\
        \vdots & \vdots & 0 & \ddots & & \vdots \\
        \vdots&\vdots&\vdots&&\hspace{0.3cm}\ddots & 0\\[0.15cm]
        -c & -c & 0 & \cdots & \hspace{0.25cm}0 & \hspace{-0.05cm}b
    \end{pmatrix}, \qquad 
    \begin{matrix}
        a=n \omega \big(\spm\coth(\omega \ell_A)+\coth(\omega \ell_C)\big)\,,\\[1ex]
        b=2 \,\omega \big(\spm\coth(\omega \ell_B)+\coth(\omega \ell_C)\big) \,, \\[1ex]
        \hspace{-2.4cm}c=\omega/\sinh(\omega \ell_C) \,.
    \end{matrix}
    \label{DihedralMatDis}
\end{equation}

\medskip
\textit{Determinant ---} We begin by reindexing the $\phi$s such that we obtain the modified matrix
\begin{equation}
    M_n^{(D)}=\begin{pmatrix}
        a & -c & -c & \cdots & -c & 0 \\[0.2cm]
        -c  & b & 0 & \cdots & 0 & -c \\
        \vdots & 0 & \hspace{0.2cm}\ddots&  & \vdots & \vdots \\
        \vdots  & \vdots &  & \hspace{0.4cm}\ddots & 0 & \vdots \\[0.1cm]
        -c  & 0 & \cdots & 0 & b & -c \\[0.1cm]
        0 & -c & -c & \cdots & -c & a
    \end{pmatrix}.
\end{equation}
By subtracting the first column by the last and the first row by the last, we can show that its determinant is
\begin{equation}
    \det M^{(D)}_n=ab^{n-1}(ab-2nc^2) \spp.
\end{equation}

\subsubsection{Adjacent regions on the boundary}

For adjacent regions on the boundary, the replica graph is represented Fig.~\ref{DihedralAdj}, and the related matrix $M^{(D)}_n$ is $2n\times 2n$ and reads
\begin{equation}
M^{(D)}_n=\begin{pmatrix}
    a & -c &  0 & \cdots & 0 & -c\\[0.5ex]
    -c & b & -c & 0 & \cdots& 0 \\
    0 & -c & \hspace{0.2cm}a & -c& & \vdots \\[0.5ex]
    \vdots & & \hspace{0.2cm}\ddots & \hspace{0.2cm}\ddots&\hspace{0.2cm}\ddots & 0 \\
    0 & \cdots & 0 & \hspace{-0.25cm}-c & \hspace{-0.25cm}a & \hspace{-0.25cm}-c \\[0.5ex]
    -c & 0 & \cdots & 0 & \hspace{-0.25cm}-c & \hspace{-0.15cm}b
\end{pmatrix}, \qquad 
    \begin{matrix}
        a=2 \omega \big(\spm\coth(\omega \ell_A)+\coth(\omega \ell_B)\big)\,,\\[1ex]
        b=2 \,\omega \big(\spm\coth(\omega \ell_B)+\coth(\omega \ell_C)\big) \,, \\[1ex]
        \hspace{-2.4cm}c=\omega/\sinh(\omega \ell_B) \,.
    \end{matrix}
\label{DihedralMatAdj}
\end{equation}

\pagebreak
\textit{Determinant ---} The determinant is computed in \cite{Berthiere:2023bwn} (A.5). We reproduce here the main steps of this calculation.
We begin by dividing every term by $c$ to simplify the matrix, and as this matrix has equal numbers of $a$ and $b$, we can replace all the diagonal terms by $\sqrt{ab}/c$. The new matrix becomes a circulant matrix
\begin{equation}
    \tilde{M}_n^{(D)} = \text{circ} \Big (\sqrt{ab}/c,-1,0,...,0,-1 \Big) \,.
\end{equation}
This matrix has determinant of the form
\begin{equation}
    \det \tilde{M}_n^{(D)} = \prod_{j=1}^{2n} \Bigg(\frac{\sqrt{ab}}{c}-e^{\frac{2\pi i j}{2n}}-e^{\frac{2\pi i (2n-1)j}{2n}}\Bigg)= \prod_{j=1}^{2n} \Bigg(-\frac{\sqrt{ab}}{c}+2\cos\frac{2\pi j}{2n}\Bigg),
\end{equation}
which can be analytically continued to non-integer $n$ by a property of Chebyshev polynomials of the first kind $T_{2n}$
\begin{equation}
    \prod_{j=1}^{2n} \bigg(2a+2\cos\frac{2\pi j}{2n}\bigg)=2(T_{2n}(a)-1) \,, \qquad T_{2n}(\cos \theta)=\cos(2n\theta).
\end{equation}
Putting everything together finally yields
\begin{equation}
    \det M^{(D)}_n = c^{2n} \det \tilde{M}_n^{(D)} =4 c^{2n} \sinh^2(n\theta) \,, \quad\theta=\text{arccosh}(\sqrt{ab}/2c)=\text{arccosh}(\sqrt{\eta^{-1}}) \,.
\end{equation}


\twocolumngrid

\let\oldaddcontentsline\addcontentsline
\renewcommand{\addcontentsline}[3]{}

\bibliographystyle{utphys} 
\bibliography{ref_GME_LFT.bib}

\let\addcontentsline\oldaddcontentsline

\end{document}